\documentclass[preprintnumbers,prd,twocolumn,showpacs,floatfix,superscriptaddress]{revtex4-1}

\usepackage{amsfonts} 
\usepackage{amssymb} 
\usepackage{amsmath} 
\usepackage{graphicx} 
\usepackage{subfigure} 
\usepackage{array} 
\usepackage{dcolumn} 
\usepackage{bm} 
\usepackage{latexsym} 
\usepackage{longtable} 
\usepackage{hyperref} 
\usepackage{bbold}
\usepackage{mathtools}
\usepackage{soul}

\usepackage[utf8]{inputenc}
\usepackage[braket]{qcircuit}
\usepackage[dvipsnames,usenames]{xcolor}
\usepackage[normalem]{ulem}

\newcommand{\UNITN}{{Dipartimento di Fisica, University of Trento, via Sommarive 14, I–38123, Povo, Trento, Italy}}
\newcommand{\TIFPA}{INFN-TIFPA Trento Institute of Fundamental Physics and Applications,  Trento, Italy}
\newcommand{\LANL}{Theoretical Division, Los Alamos National Laboratory, Los Alamos, New Mexico, 87545}

\begin{document}
\preprint{LA-UR-25-31666}
\title{ Neutrino Flavor Evolution in High Flux Astrophysical Environments}
\author{J. Carlson}
\affiliation{\LANL}
\author{A. Roggero}
\affiliation{\UNITN}
\affiliation{\TIFPA}
\author{Duff Neill}
\affiliation{\LANL}

\date{\today}
\begin{abstract}
We examine neutrino evolution in astrophysical environments where the neutrino flux is very large, including core-collapse supernovae and neutron star mergers. In these environments, the neutrino-neutrino and neutrino-antineutrino interactions are crucial.  We include non-forward scattering of neutrinos and anti-neutrinos in a semi-classical treatment.  Because of the large scale of neutrino momenta (2-10 MeV), the quantum evolution  problem can be treated as a 
sum over incoherent paths in the 
and flavor of each neutrino. The phases between different neutrinos are essentially random because of the large kinetic terms. Momentum is conserved at each vertex, and important flavor symmetries are retained.  
Dynamics in the many-body neutrino system enable rapid equilibration in the energy and angular distributions of all flavors,
and an equilibration of products of neutrino and anti-neutrino densities for each flavor at either large or zero background
matter density.
We also describe the evolution at moderate densities where the mass eigenstates differ for
neutrinos and antineutrinos, and with time-varying background matter densities.
The evolution maintains relevant symmetries and reduces to standard MSW oscillations in the appropriate limits.
The rapid equilibration in energy and flavor can significantly impact energy deposition and nucleosynthesis in high-flux astrophysical environments, and potentially flavor energy relations in terrestrial supernovae neutrino observations.
\end{abstract}

\maketitle

\section {Introduction}

For many astrophysical environments like the sun, neutrino flavor evolution is well-understood through single-particle
evolution and the MSW mechanism~\cite{PhysRevD.17.2369,MIKHEYEV198941}.  In environments with large neutrino fluxes, however, neutrino-neutrino and neutrino-antineutrino
interactions can play a role~\cite{pantaleone1992neutrino,PhysRevD.46.510,Qian:1994wh,pastor2002physics,Duan:2005cp} (see also Refs.~\cite{annurev:/content/journals/10.1146/annurev.nucl.012809.104524,CHAKRABORTY2016366,RevModPhys.96.025004} for recent reviews).   Here we consider the evolution of neutrino flavor, including
kinetic terms (neutrino momenta), the vacuum plus matter (MSW) one-body interactions, and 
two-neutrino interactions. A complete quantum treatment of this problem for large systems of neutrinos
and antineutrinos is prohibitive due to the large number of quantum amplitudes involved and the long time evolution required. This major challenge has motivated efforts to explore the use of approximate many-body techniques~\cite{Bell:2003mg, Friedland:2003dv, Friedland:2003eh, Friedland:2006ke, McKellar:2009py, Balantekin:2006tg, Pehlivan:2011hp, Pehlivan:2014zua, Birol:2018qhx, Patwardhan:2019zta, Cervia:2019res, Rrapaj:2019pxz, Roggero2021a, Roggero2021b, Xiong:2021evk, Martin:2021bri, Patwardhan:2021rej, Roggero2022, Cervia:2022pro, Illa:2022zgu, Lacroix:2022krq,Siwach:2022xhx, PhysRevC.110.045801, Martin:2023gbo, Martin:2023ljq,PhysRevD.109.103037,PhysRevD.110.103027,Cirigliano:2024,d6w8-7j9s,PhysRevResearch.7.023228,manginbrinet2025threeflavorneutrinooscillationsusing} and, in recent years, also quantum computing~\cite{Hall:2021rbv,Yeter-Aydeniz:2021olz,PhysRevA.106.052605,Amitrano2023,Siwach:2023wzy,Turro:2025,PhysRevD.111.043017,gjr1-lf8s,kiss2025neutrinothermalizationrandomizationquantum}

In this work, we develop a semiclassical treatment analogous to standard classical mechanics but including the time evolution
of the momenta and flavor amplitudes for each neutrino. 
The many-body quantum system equilibrates at a similar rate to the forward scattering Hamiltonian,
as discussed below.
Though the actual final state is highly entangled, simple one body observables
can be accurately evaluated as averages over trajectories in this semi-classical approach.

\section{Hamiltonian}

  We consider a system of $N$ neutrinos at a finite density described by the Hamiltonian of the form
\begin{equation}
	H = H_0 +  H_{1,v} + H_{1,m} + H_2\;.
\end{equation}
 The Hamiltonian is written in the basis of products of single-particle states, where each neutrino has a definite flavor state $\chi_i$ 
and a definite momentum ${ \bf k}_i$. 

The first term $H_0$ is the leading order kinetic term 
\begin{equation}
H_0 = \sum_{i=1}^N  \ |{\bf k_{i}}|
\end{equation}
for massless neutrinos, 
where $i$ runs over the $N$ neutrinos. 
The $H_1$ terms are the vacuum ($H_{1,v}$) and matter terms ($H_{1,m}$)  
well-studied for solar neutrinos and extraterrestrial 
neutrinos passing through the earth, for example.
They are also being studied in large experimental campaigns using
accelerator neutrinos to determine the mass hierarchy and potential 
CP violation in the neutrino sector. Finally, the last term $H_2$ describes the neutral-current mediated interaction between neutrinos.

In this paper, we consider Dirac neutrinos with a normal mass ordering  and no CP violation.
Extensions to Majorana neutrinos with CP violation are an intriguing  subject for future study.
With these choices, the numbers of neutrinos and antineutrinos are individually conserved.

The two-neutrino interactions are flavor operators
times a vertex function of momenta nearly independent of the magnitude of momentum transfer ${\bf q}$ (since it is
a very short-range interaction) times a function of the angles of incoming and outgoing momenta.
The interactions operate on the entire flavor space of each neutrino, with a constant term
plus a flavor exchange term (for neutrino-neutrino interactions) or a constant plus a term coupling a neutrino-antineutrino pair with the same initial flavor to final states of the other flavors. 
This is the same Hamiltonian studied in Ref.~\cite{Cirigliano:2024}; additionally, we use semiclassical
methods to address a large number of neutrinos, study cases with different background matter profiles, and address the impact of a mixture of neutrinos and antineutrinos.

We express a general neutrino state with a fixed momentum in a coupled flavor-momentum basis as follows
\begin{equation}
\label{eq:sp_nu}
\rvert \psi_m(i) \rangle= \sum_{f=1}^{N_f} \chi_{mf}(i) \rvert f\rangle\otimes\rvert{\bf k}_m \rangle\;,
\end{equation}
where states $\ket{f}$ form a basis for the $SU(N_f)$ flavor space (i.e., for $N_f=3$ these are $\ket{\nu_e}$, $\ket{\nu_\mu}$, and $\ket{\nu_\tau}$). The coefficient matrix $\chi_{mf}(i)$ stores the flavor amplitudes of the single particle state. For anti-neutrinos, we similarly write
\begin{equation}
\label{eq:sp_nubar}
\rvert \overline{\psi}_m(i) \rangle= \sum_{f=1}^N \overline{\chi}_{mf}(i) \rvert \overline{f}\rangle\otimes\rvert{\bf k}_m \rangle\;.
\end{equation}
A general many-body state with $N_\nu$ neutrinos and $N_{\overline{\nu}}$ anti-neutrinos can be expressed as a linear combination
\begin{equation}
   \rvert \Psi\rangle = \sum_n  \alpha_n \rvert \phi_n\rangle\;,
\end{equation}
where the states $\rvert \phi_n\rangle$ are product states of the form
\begin{equation}
\label{eq:prod_states}
\rvert \phi_n\rangle= \bigotimes_{i=1}^{N_\nu} \rvert \psi_{m_n}(i) \rangle\;
	\bigotimes_{j=1}^{N_{\overline{\nu}}} \rvert \overline{\psi}_{\overline{m}_n}(j) \rangle\;,
\end{equation}
where the labels $m_n$ and $\overline{m}_n$ for individual neutrinos and antineutrinos are chosen to avoid double occupancy of any specific single-particle state. The average energy of individual neutrinos is on the order of $2-10$ MeV; the density outside the neutrino-sphere is too small to reach even a quasi-degenerate regime (see also Ref.~\cite{Martin:2023}). For this reason, they can be regarded as essentially distinguishable (or Boltzmann) particles.

From the definitions in Eq.~\eqref{eq:sp_nu} and Eq.~\eqref{eq:sp_nubar} we see that each particle can be in a linear combination of $N_f$ flavor states and a specific momentum mode. This implies that the many-body states $\rvert \phi_n\rangle$ introduced above have a well defined momentum and are eigenstates of the flavor-independent kinetic term $H_0$ in the Hamiltonian.
One-body terms in the Hamiltonian $H$ can be expressed more explicitly as
\begin{equation}
H_{1,v} = \sum_{i=1}^N H^{v}_{1,i}\quad H_{1,m} = \sum_{i=1}^N H^{m}_{1,i}\;.
\end{equation}
The first term describes vacuum oscillations and $H^{v}_{1,i}$ is the standard vacuum Hamiltonian containing the experimentally constrained mixing angles and mass squared differences:
\begin{equation}
	H^{v}_{1,i} =U_{f}\begin{pmatrix}
	\frac{m_1^2}{2| {\bf k}_i|}&0&0\\
	0&\frac{m_2^2}{2| {\bf k}_i|}&0\\
	0&0&\frac{m_3^2}{2| {\bf k}_i|}\\
	\end{pmatrix}U^\dagger_{f}\;,
\end{equation}
where $U_f$ is the basis transformation from flavor to mass (for 3 flavors, it is the PMNS matrix~\cite{Pontecorvo:1957cp,10.1143/PTP.28.870}) and $m^2_j$ are the squared masses of the energy eigenstates.

The neutrino matter coupling is described by the standard  diagonal one-body
potential arising from forward scattering from matter (assuming zero density of $\tau$ leptons):
\begin{equation}
	H^m_{1,i} =  \pm \sqrt{2} G_F  (\Pi_{e} \; \rho_e 
		     + \Pi_{\mu} \; \rho_{\mu} )
\end{equation}    
where $\Pi_x$ are projections on $x$ flavor states; $\rho_e$ ($\rho_{\bar{e}}$) and $\rho_\mu$ ($\rho_{\bar{\mu}}$) are the respective electron (positron) and muon (anti-muon)
densities, while $G_F$ is Fermi's constant.
The overall sign $\pm$ 
is for neutrinos and anti-neutrinos, respectively.

In astrophysical environments like core-collapse supernovae, the flux of neutrinos and antineutrinos is so large that the neutrino-neutrino coupling becomes significant~\cite{pantaleone1992neutrino,PhysRevD.46.510}. The forward scattering neutrino-neutrino interaction between many-body states with fixed momenta for each neutrino takes the following form:
\begin{equation}
	H^{fs}_{2,\nu\nu}\left(\{{\bf k}_i\}\right) = \frac{\sqrt{2} G_F\rho_\nu}{2N_\nu}    \sum_{i<j} \left(1-\hat{\bf k}_i\cdot\hat{\bf k}_j\right)\vec{\lambda}_i \cdot \vec{\lambda}_j,
\end{equation}
where the $N_f$ component vectors of operators $\vec{\lambda}_i$ are the Pauli ($N_f=2$) or Gell-Mann ($N_f=3$)
matrices acting on the different flavor amplitudes of each neutrino.
The neutrino density is $\rho_\nu=N_\nu/V$, so the interaction can be written as independent of neutrino number for a fixed volume. An equivalent but more convenient parametrization of this interaction relies on the observation that
\begin{equation}
\vec{\lambda}_i \cdot \vec{\lambda}_j = 2P_{ij}-\frac{2}{N_f}\;,
\end{equation}
where the $P_{ij}$ is a flavor permutation operator exchanging each amplitude of one spinor with the other. Since the constant term does not affect time evolution, we can write for any $N_f$ the forward scattering interaction as
\begin{equation}
	H^{fs}_{2,\nu\nu}\left(\{{\bf k}_i\}\right) = \frac{\sqrt{2} G_F\rho_\nu}{N_\nu}    \sum_{i<j} \left(1-\hat{\bf k}_i\cdot\hat{\bf k}_j\right)P_{ij}\;.
\end{equation}
 For the anti-neutrinos, the forward scattering contribution has a similar structure
\begin{equation}
	H^{fs}_{2,\bar{\nu}\bar{\nu}}\left(\{{\bf k}_i\}\right) = \frac{\sqrt{2} G_F\rho_{\bar{\nu}}}{N_{\bar{\nu}}}    \sum_{i<j}^{N_{\bar{\nu}}}\left(1-\hat{\bf k}_i\cdot\hat{\bf k}_j\right)P_{ij}\;,
\end{equation}
while for the mixed neutrino-anti-neutrino term, we have
\begin{equation}
\label{eq:nunubar_int}
	H^{fs}_{2,\nu\bar{\nu}}\left(\{{\bf k}_i\}\right) = \frac{\sqrt{2} G_F\rho_{\bar{\nu}}}{N_{\bar{\nu}}}    \sum_{i=1}^{N_{\nu}}\sum_{j=1}^{N_{\bar{\nu}}}\left(1-\hat{\bf k}_i\cdot\hat{\bf k}_j\right)Q_{ij}\;,
\end{equation}
where we used $\rho_\nu/N_\nu=\rho_{\bar{\nu}}/N_{\bar{\nu}}=1/V$. The operators $Q_{ij}$ act on the flavor spaces of particles $i$ and $j$ as
\begin{equation}
Q_{ij} = \sum_{f\neq f' } \rvert f,\overline{f}\rangle\langle f',\overline{f'}\lvert\;.
\end{equation}

The total energy is proportional to the number of neutrinos or the volume at fixed density, the same as the one-body terms.
The neutrino-neutrino two-body interactions are representation-independent for any orthonormal single-particle basis.  It is the same in either the vacuum basis or the matter basis.
The same is true for the neutrino-antineutrino interaction when the matter potential is either zero or large and the flavor basis is essentially the same as the matter basis. The general case is discussed later in Sec.~\ref{sec:mod_matter_dens}.

Finally, the full interaction term including momentum exchanges can be written as
\begin{equation}
	H^{fs}_{2,\nu\nu}\left(\{{\bf k}_i\}\right) = \frac{\sqrt{2} G_F\rho_\nu}{N_\nu}   \int d^3{\bf q} \sum_{i<j}O_{ij}({\bf q})\;,
\end{equation}
where the operator $O_{ij}({\bf q})$ acts on both the flavor and momentum space of particles $i$ and $j$ as
\begin{equation}
O_{ij}({\bf q})= h({\bf q},{\bf k}_i,{\bf k}_j)\frac{1+P_{ij}}{2}\rvert{\bf k}_i+{\bf q},{\bf k}_j-{\bf q}\rangle\langle{\bf k}_i,{\bf k}_j\lvert\;,
\end{equation}
while we introduce, following Ref.~\cite{Cirigliano:2024}, the function
\begin{equation}
 h({\bf q},{\bf k}_i,{\bf k}_j) = f^*({\bf k}_i+{\bf q},{\bf k}_j-{\bf q})f({\bf k}_i,{\bf k}_j)\;,
\end{equation}
where we have defined (see~\cite{Cirigliano:2024})
\begin{equation}
\begin{split}
f({\bf p},{\bf q})=\sqrt{2}\bigg(&e^{-i\phi_{\bf p}}\sin\left(\frac{\theta_{\bf p}}{2}\right)\cos\left(\frac{\theta_{\bf q}}{2}\right)\\
&-e^{-i\phi_{\bf q}}\sin\left(\frac{\theta_{\bf q}}{2}\right)\cos\left(\frac{\theta_{\bf p}}{2}\right)\bigg)\;.
\end{split}
\end{equation}

Inclusion of the non-forward scattering is important because in an incoherent sum over paths
the coupling between neutrino's flavor in essence changes with time.

\subsection{Time Evolution and Energy Scales}
\label{ssec:scales}

To create  a semiclassical approach to the time evolution it is useful to compare the energy scales of the various terms in the Hamiltonian. 
These scales control oscillation lengths and potential interference in the path integral 
describing time evolution.

The largest scale by far is the kinetic term $H_0$ since, as mentioned above, astrophysical neutrinos in these settings have energies of order $2-10$ MeV. The associated oscillation
lengths are $\hbar c/ E$ or $20-100$ Fermi.  These oscillation lengths are of order $10^{-10}$ or smaller than the other oscillation lengths. This has important consequences and, as we discuss further below, suggests that a semiclassical treatment might be appropriate.

On the other end of the scale we have the vacuum oscillation energies.
These are of order $\delta m^2 / E  \approx 10^{-4} eV^2 / 10$ MeV
or $10^{-11} $ eV. These scales are of order 10 km for $\delta m_{23}^2$
and a 10 MeV neutrino, or even larger for the $\delta m_{12}^2/E$ mass scale.
A substantial background matter component can, however, dramatically shorten
these oscillation lengths. For example, at 
electron densities of order $10^{38} / cm^3$, the oscillation length
for the electron flavor is of order one meter. This can be comparable
to the scale of the neutrino-neutrino interaction in dense astrophysical environments.

In the region where neutrinos are dense but outside the proto-neutron
star core in a supernova, the two-body interactions have typical length scales ranging from
centimeters to meters.  In Ref.~\cite{Martin:2023}
we give an estimate of 1 cm at
a radius of 50 km.  At larger distances from the core, the density
decreases like $1/R^2$, with an additional reduction from the more-forward
scattering of up to another $1/R^2$.  Non-forward scattering might
slow the decrease of the interaction with distance to some degree.
Forward scattering would give a length of approximately
1 meter at a radius of 150 km, still dramatically longer than the $H_0$
scale but much smaller than the scale of the vacuum 
term.  
At 500 km, the distance would be roughly 100 meters, still short 
compared to vacuum oscillation lengths, but perhaps comparable to
vacuum plus matter.

\section{Time evolution and kinetic decoherence }

The time evolution of the full system is governed by 
\begin{equation}
	\rvert \Psi(t) \rangle = \exp[ - i H t] \  \rvert \Psi(0) \rangle
\end{equation}
where $\rvert \Psi(0) \rangle$ the initial state which we assume to be a single product state of the form in Eq.~\eqref{eq:prod_states} above. A more general incoherent sum of product states can be treated equivalently.  

The time evolution is unitary, conserving the norm of the state.
For this Hamiltonian the total number of neutrinos and anti-neutrinos are separately conserved.
These can change from other
processes like capture and emission that are not considered here. We do expect however that these processes will only enhance
the decoherence.

The time evolution of the system can be written as a path integral
with one and two-body interaction terms. Since the one-body term can be solved explicitly,
we rewrite it in a form where the one-body terms are explicitly included
and insert two-body vertices at arbitrary times:
\begin{eqnarray}
	e^{- i H t} & = & \sum_{n = 0}^{\infty}   (-i)^n \nonumber  \nonumber \\
	& &  \mathcal{T}	\int d t_n \int dt_{n-1} ...  \int d t_1 \int d t_0   
\sum_{i_n,j_n,,...,i_1,j_1} \nonumber \\
	&	&\exp[ - i H_{01} t_n] H_{2;i_n,j_n}  \exp [ -i H_{01} t_{n-1} ]
	  ... \nonumber \\ 
	& &  \exp [ - i H_{01} t_1]  H_{2;i_1,j_1} \exp[-i H_{01} t_0] ,
\label{eq:prop}
\end{eqnarray}
with the requirement that the total time $t = \sum_i t_i$. 
In this expression  we have used $H_{01} = H_0 + H_{1,v} + H_{1,m}$, the sum of the
kinetic, vacuum and mass one-body terms, while $H_2$ is the sum of the neutrino-neutrino, antineutrino-antineutrino and neutrino-antineutrino terms. Our approach then relies on sampling both the sum and the integral in the same spirit of Stochastic Series Expansion versions of the Quantum Monte Carlo method~\cite{PhysRevB.26.5033,PhysRevB.43.5950}.

The one-body evolution through $H_{01}$ is simple: 
an initial product state of neutrinos retains that character with phases
from the vacuum plus matter Hamiltonian $H_{01}$.
For convenience we use a local mass basis for the one-body evolution.
A Trotter like evolution where the one-body term is diagonal between
each two-neutrino vertex is appropriate.  If the neutrino flux is very low,
as in the Sun, one has to include an explicit time-dependence to account 
for the spatially varying matter potential. In this way the standard
MSW limit is recovered in the limit of low neutrino density.
Following the discussion in Sec.~\ref{ssec:scales}, in a supernova the time scale of two neutrino interactions is shorter than 
that due to the
changes in the mass density and we then take the $H_{1,m}$ Hamiltonian independent of time.

\begin{figure}
\includegraphics[width=3.5in]{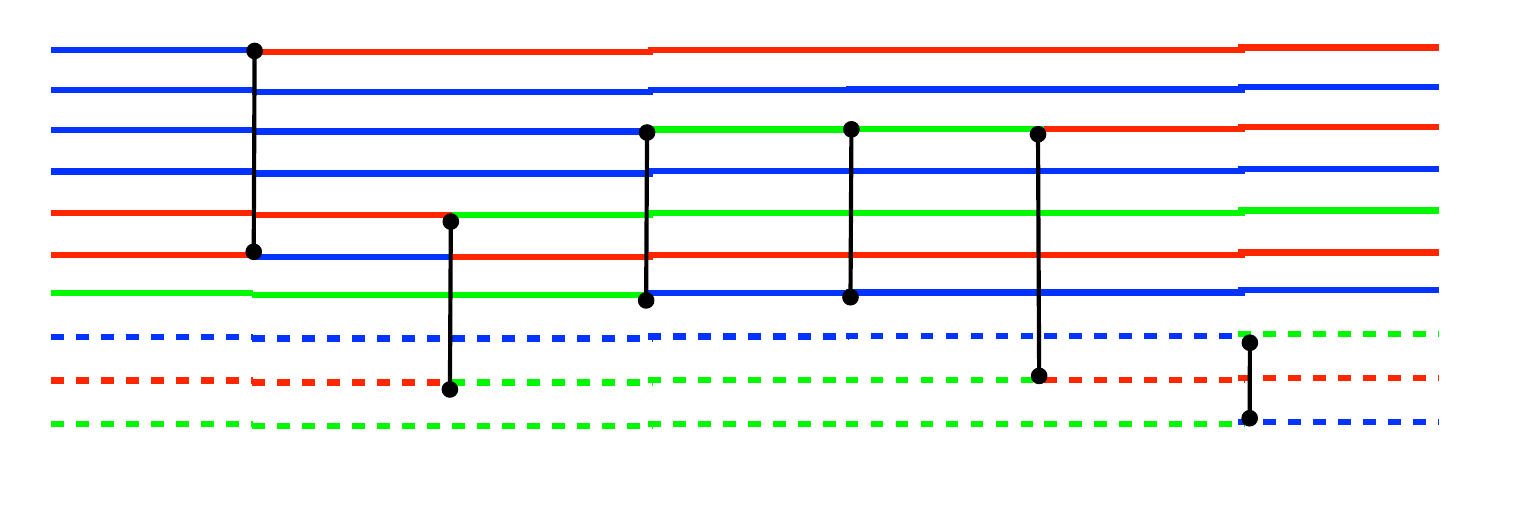}
\caption{Pictorial representation of the propagation represented by 
	Eq. \ref{eq:prop}.   Horizontal lines are neutrinos, dashed lines are anti-neutrinos, and color represents flavor. Vertical lines represent two-neutrino vertices. Only flavor evolution is indicated here, though the momenta are also changing at each vertex.}
\label{fig:pathint}
\end{figure}

The two-neutrino interaction involves flavor and momentum 
amplitudes of two neutrinos at
a time.  In this equation we write $H_2 = \sum_{i<j} H_{2,ij}$ and
sum over (or sample) individual pair operators in the path integral.
Pictorially  this can be described as shown in Fig.~\ref{fig:pathint}.

In this figure, different mass eigenstates or flavors are represented by
the colors. The horizontal lines represent diagonal time evolution in the instantaneous
mass basis between two-nucleon vertices.  The vertical lines represent two-neutrino or neutrino-antineutrino interactions, which may change the flavors.
The momentum directions can change at each vertex, but these changes are
not indicated in the figure.  The magnitude of the momenta can also change.

\begin{figure}
\includegraphics[width=3.5in]{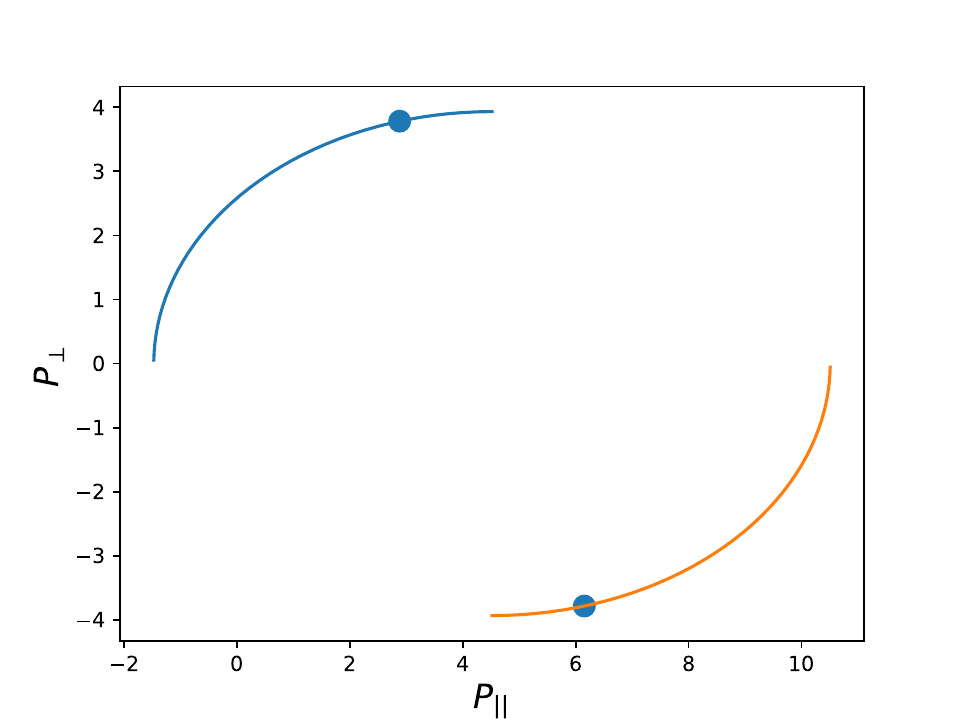}
\caption{Possible two neutrino final states conserving 
momentum and energy. Sample initial states
momenta parallel and perpendicular to the total pair momentum are shown as circles.  Possible two-neutrino final states conserving energy and momentum are shown as lines.  The two final state neutrino each lie on one of the two lines. There is an additional azimuthal symmetry as well.\label{fig:vertices-mom}}

\end{figure}

\begin{figure}
\includegraphics[width=3.5in]{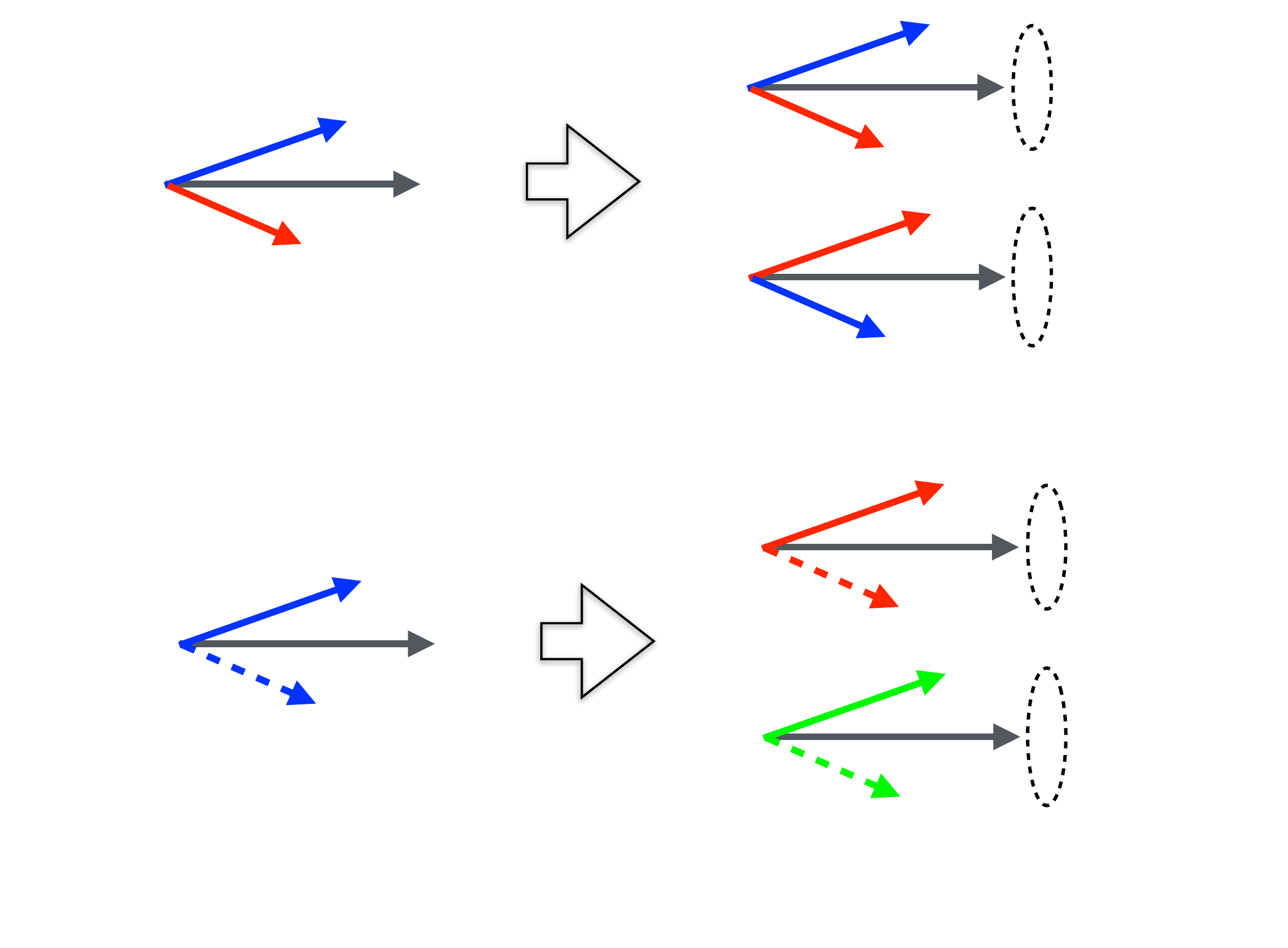}

\caption{Possible initial and final states in neutrino-neutrino (upper) and neutrino-antineutrino (lower) vertices,
indicating azimuthal rotations and possible flavor changes. The sum of the two momenta and the individual magnitudes (kinetic energy) are conserved. The final states can either exchange the flavors in the upper panel or not.  In the lower panel the 
momenta magnitudes can change (or not) between neutrino and antineutrino and the flavor of the pair can change. The dashed ovals are meant to indicate the
possible azimuthal rotation of momenta in the final state, still preserving the total momenta.}
\label{fig:vertices-flavor}
\end{figure}

In two-neutrino scattering, the final states must conserve energy and momentum.  Momentum is conserved by construction for an interaction that is a function of 3-momentum transfer. Energy conservation introduces a delta function in final minus initial state energy, reducing
the dimensions of possible momentum transfer $\mathbf{q}$ to two, as shown in Ref. \citenum{Cirigliano:2024} in the CM frame.
Figure \ref{fig:vertices-mom} shows possible final states in the lab frame
for a specific set of initial momenta, projected into components parallel and perpendicular to the total momentum of the pair. There is also an azimuthal rotation possible for each pair and different possible flavor states, see 
Fig. \ref{fig:vertices-flavor}.
The flavors can change; a neutrino entering with a specific flavor and momentum can change either momentum or flavor, or both. The result is a two-dimensional surface of final state momenta that conserve both energy and momenta with potentially more than one flavor state for each set of momenta.

The energy in the flavor degrees of freedom is
very small compared to $H_0$, and in two-neutrino scattering, its energy can vary freely, with only the total energy being conserved through
tiny variations in the magnitude of the momenta. The time-varying couplings from the scattering allow the flavor 
component of the energy to vary freely over the full many-body range.

It is valuable to compare full many-body evolution to a classical treatment where energy is exactly conserved at each scattering.
One step beyond the classical treatment is to consider the evolution of a system including one initial particle in a bath of background neutrinos
that are already  equilibrated.   This evolution is essentially what happens when a new neutrino is created in a background of N neutrinos, 
perhaps at high energy. The time evolution of this additional neutrino can be treated in linear response because one scattering of the ``impurity'' is incoherent from the next one.  This happens
because there are many more pairs of background neutrinos interacting at all times, and only
one impurity neutrino. An example of this type of physics  is  an impurity ion created in an 
electron plasma in thermal equilibrium\cite{Fetter-Walecka}.   This goes beyond a classical treatment in that it includes virtual excitations of the new neutrino
with all neutrinos in the bath, including possible momentum and energy exchanges. 
The total energy is conserved, but the division between an impurity and the background changes with time.
Linear response treats the interaction of the background neutrinos to all orders but considers
only one scattering of the new neutrino with the many-body system.  As long as successive interactions of the original neutrino are incoherent,
linear response should be sufficient.  The evolution of this impurity neutrino is governed by 
$ \int d{\bf q} \int d \omega \omega S(q,\omega)$, with limits of integration determined by the initial momentum and energy of the ``impurity''.
The average time required for one scattering is determined by $t$ times the sum of the magnitudes of all transition matrix elements from the initial state and is
hence proportional to $1/G_F$. In a single scattering, the impurity will on average lose half of its energy since it can exchange with another neutrino.
A few such scatterings will be sufficient to reach equilibrium.  Detailed kinematics and their impact should be studied in more detail.

In the full many-body quantum evolution, virtual excitations are also crucial. Only after the full propagation over a time $t$, where all pair interactions are included, 
is energy conservation imposed.   This gives potentially even more freedom than evolution with one impurity.  The full evolution involves
the time evolution of all particles simultaneously, so we expect a similar amount of time to be required. As in standard evolution with sufficient random coupling, the equilibration time
will be
governed by the variance: the expectation value of $\langle H^2 \rangle $ minus the expectation value of $\langle H \rangle$ squared.
In a product state of spins, timed with plane waves in space, only the sum over spin terms for each pair contributes to the variance.  
The resulting equilibration time $t$ is of
the order of  the inverse of the variance per particle~\cite{MARGOLUS1998188,Deffner_2017,PhysRevLett.120.060409}
\begin{equation}
t \approx  \pi / \sqrt{   \langle (H_2/N)^2 \rangle - \langle H_2/N \rangle^2}.
\end{equation}
again proportional to $G_F$. The factors of 1/N are for extensive observables, the minimum average time for any exchange in the system is a factor of N faster. For an initial product state and a time-independent Hamiltonian the variance is $O(N)$ which implies this equilibration time is $O(\sqrt{N})$ (see also~\cite{Martin:2023,PhysRevResearch.7.023157}). Note that this is a lower limit
on the equilibration time resulting from assuming no interference.

In addition we can estimate the increase in equilibration time from enforcing energy conservation in the kinetic term at each vertex.  The result is a two-dimensional surface at each vertex, as for the two-body case.  This only increases the equilibration time by only a modest factor of 3-4, obtained by estimating the ratio of states without energy conservation as compared to with energy conservation at each vertex,
and then equating the number of states reached in the two cases for different finite times
(see Eq. \ref{eq:finalstates}). This insensitivity is due to the exponential growth of the number of non-zero amplitudes with time, as discussed below.

The difference from the forward-scatting case is that the implicit spin Hamiltonian here is time-dependent so what matters is the average variance over time. We can get a constant time scale  if the variance becomes proportional to $N^2$ on a timescale that does not depend on $N$, as is expected for an auxiliary field treatment of short-range interactions. The variance is basically given by the sum over all pairs and
all momentum transfers from the initial state to all one pair momentum transfers. The variance here is somewhat larger than for forward scattering calculations, hence the time scale is similar or shorter. For forward scattering only terms that exchange spins enter the variance. Those same terms enter here along with contributions from
terms diagonal in the spin but off diagonal in momenta. The 
spin exchange terms are the same in the two calculations, but in the full angular dependent coupling the amplitudes
are spread over many incoherent momentum modes. This time scale is considerably shorter than
the standard distance scale of the inverse of the two-body cross section times the density obtained
in purely classical calculations, as we show in more detail in Appendix~\ref{app:twob_vertex}. The rapid equilibration depends upon the short-range nature of the interaction and could be tested in cold atom experiments~\cite{coldatomspeed:2025}.
In such experiments the two-body interaction can be rapidly increased from essentially non-interacting to a finite scattering length, we expect equilibration times for low to modest interactions to scale with the inverse of the scattering length.

The dephasing effect caused by the large kinetic energy scale is often called 'kinetic decoherence'. As it arises from the flavor independent kinetic terms, it essentially implies that different paths with different momenta for the neutrinos do not interfere.
The spatial part (momentum directions and amplitudes for each neutrino) can essentially be treated classically; however, in a full calculation, energy conservation must
be considered for the initial and final states.  In the quantum time evolution, this yields a random overall phase for each neutrino as a function of time, with a constraint only on their sum.
Enforcing energy conservation at each vertex implies only a single overall phase for all the momentum degrees of freedom, which does not impact observables.

In principle, for the same kinematic (momentum) paths, the relative phases
of different neutrinos could still interfere.  
The flavor state of the system is essentially described by permutations of the original flavors plus conversions of neutrino-antineutrino pairs of one flavor to another. Since it is an all-to-all interaction, one pair exchange is rapidly
followed by another between a different pair, leading to an entangled many-body state which, however, locally (i.e., for individual neutrinos) will look indistinguishable from an incoherent flavor state. In the interior, where the eigenstates are split by larger effective $\delta m^2$, this also contributes to flavor decoherence.  

Potential interference between different flavor amplitudes of individual neutrinos remains. This is the interference that gives rise to the MSW effect,
and should remain in the limit of low neutrino density.
Using the fact that different spatial paths do not interfere, we can make predictions about the time-dependent
 expectation values of single neutrino observables without 
keeping track of all the phases that would be required to 
reverse the unitary time evolution and return to the initial state.
This is the essence of equilibration or thermalization in quantum evolution discussed in previous publications~\cite{Martin:2023}.
Here, the neutrinos equilibrate very rapidly, subject to symmetries
including energy and momentum conservation, as well as conservation of the total
number of neutrinos and the total number of antineutrinos.

\section{Semiclassical Approximations}

\begin{figure}
\includegraphics[width=3.5in]{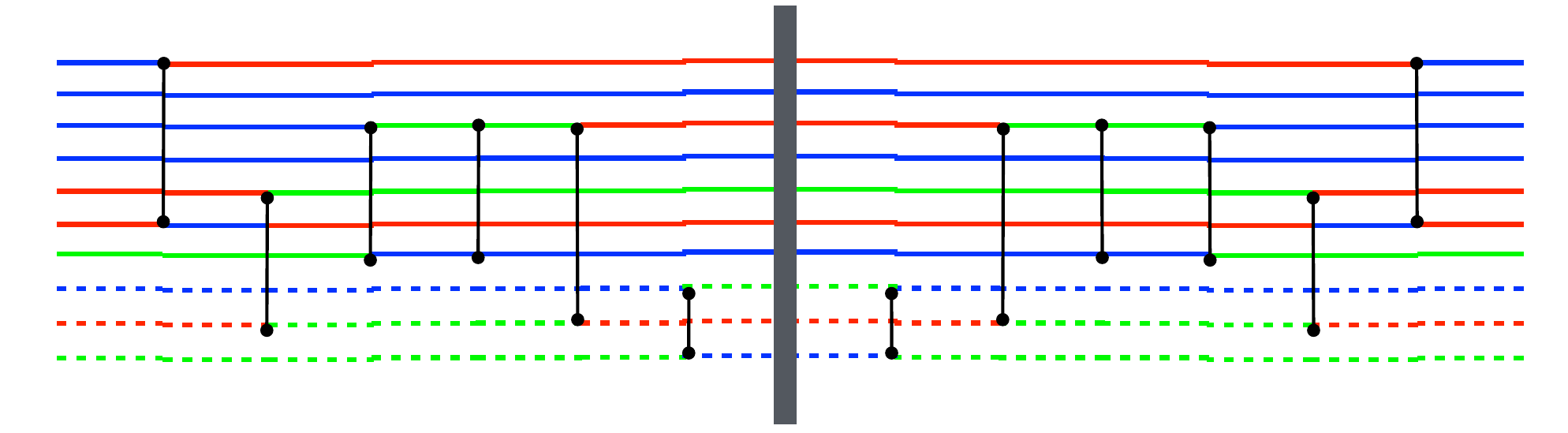}
\caption{Pictorial representation of the calculations of observables
at the black vertical bar from the squares of the relevant amplitudes.
The path integral on the left and right is the same with phase information
remaining from flavor amplitudes within each neutrino.}
\label{fig:ampsquared}
\end{figure}

To compute semiclassical approximations to one-body observables as a function of time, we take advantage of the rapid decoherence in two- or more-body
neutrino amplitudes to write them as a sum over incoherent paths defined in the product space.
The times between vertices are governed by the inverse of the sum of the magnitudes of the two-body off-diagonal matrix elements; the
sampling of specific off-diagonal transitions is taken from the scaled sum of the squares of various possible off-diagonal matrix elements.
Since the different paths are nearly incoherent and each contributes a positive contribution
to the probability, 
Monte Carlo methods can be used to sample the times and the transitions. 
This is represented for a single path in Fig. \ref{fig:ampsquared}.
The vertical bar represents the time where expectation values are taken.

For constant background densities of leptons, the one-body interaction is
constant in time. In this case, we take the initial state to be a 
product state of neutrinos in the mass basis. In this basis
, the $H_{0} + H_{1}$ is diagonal and can be represented, up to a phase, 
by specific mass states
invariant in time (represented by a single color connecting each two-neutrino
vertex). The random exchanges of flavor and momenta
give rise to equilibration of the energy of the
different flavors of neutrinos. The rapid
random conversion of $\nu {\bar \nu}$ pairs of a given flavor
or mass state
to another flavor or mass state gives rise to an equilibrium condition:
\begin{equation}
	\rho( \nu_e) \rho ({\bar \nu}_e) =
	\rho( \nu_\mu) \rho ({\bar \nu}_\mu) =
	\rho( \nu_\tau) \rho ({\bar \nu}_\tau) 
\end{equation}
for a given initial difference between the numbers of neutrinos and the numbers of anti-neutrinos in each mass eigenstate:
\begin{equation}
\Delta_{\nu\bar{\nu}}=\rho_\nu-\rho_{\bar{\nu}}\;.
\end{equation}
 
This difference remains fixed for Dirac neutrinos in a constant background matter density.  The equilibration is simplest in mass eigenstates, 
but at large electron and muon densities, the flavor and mass eigenstates 
coincide, resulting in equilibration in both mass and flavor. This equilibrium will hold in any environment
where the mass eigenstates of neutrinos and antineutrinos coincide. 
These results have been discussed extensively in \cite{Martin:2023}.

With the semiclassical picture, we can also explore the time evolution 
in a spatially varying background matter
density or for smaller finite constant background densities, where
the mass eigenstates of neutrinos and antineutrinos do not coincide.
Previously, the equilibrium distribution did not depend upon 
microscopic details other than the symmetries of the Hamiltonian.
Here, the time or distance scale of the two-neutrino interactions versus
variation in background density can play an important role.

To evaluate observables as illustrated in Fig.~\ref{fig:ampsquared},
we need to average over paths.  Starting from a product state, one computes
the non-zero matrix elements $H_{ij}$ to all possible future states $j$.
The Hamiltonian is a sum of pair terms, so for each pair, we can calculate the
sum of the magnitudes of the off-diagonal matrix elements. The time for this vertex
can be obtained by sampling from an exponential distribution with time proportional to the
inverse of this sum.  This represents the time to evolve into any orthogonal state.
To evolve, one picks the shortest time from those obtained for all pairs.

The particular final state chosen is taken from the squares of the transition amplitudes,
normalized to one for all transitions.  In this manner, one is incorporating unitarity, the correct 
time dependence, and the probability of landing in a particular state.
At this point, we have a new product state and can repeat the process.  All the previous pair times
from pairs not involving either of the chosen pair are retained, while those involving either of the
chosen pair are resampled from the new off-diagonal matrix elements. This method retains independent 
evolution of clusters of neutrinos that do not interact.

For a concrete example, assume we start with an initial product state. 
For a given pair $(i,j)$ at short times $\delta t$ the transition probability can be approximated as
\begin{equation}
\begin{split}
	T_{ij}(\delta t) =\!\! &\sum_{k\neq m_i,l\neq m_j}\!\! | \langle \psi_{k},\psi_{l} \lvert  \delta t H_{2;ij} 
		\rvert\psi_{m_i},\psi_{m_j} \rangle |\;,\\  
\end{split}
\end{equation}
where the sum is over possible orthogonal final states with momenta and flavors. This expression is valid for short times $\delta t$, for an isolated
pair. 
The amplitude of each state orthogonal to the initial state grows linearly in time:
\begin{equation}
	  \alpha_j (\delta t) = V_{ij} (\delta t)\;,
\label{eq:nexttime}
\end{equation}
where $i$ is the initial state and $V_{ij}$ is the transition matrix element of the Hamiltonian.
In our semiclassical algorithm, the extremely short-time behavior can be better approximated by taking the time as the larger of two times sampled from above, since the expectation value
changes only when both the left- and right-hand sides of the propagation include a transition.

Note that we can change the time propagation to $\exp [-i (H-E_0) t]$ without changing
the observables. Here we take $E_0$ to be the sum of the kinetic energies of the
individual neutrinos.  For energy conserving time propagation we can split this into
a sum of phases for each neutrino so mass eigenstates do not precess. Any state with the
same total kinetic energy, arbitrarily divided into energies for each neutrino, can
be treated this way.

At very short times the amplitude for off-diagonal transitions is linear in time
and proportional to $G_F t$, so the probability of these new states is proportional to
$(G_F t)^2$. The total transition probability for an out-of-equilibrium neutrino impinging
on an eigenstate, thermal state or any time-independent background state will then reduce to a sum of incoherent cross sections for each possible transition.

For larger times the number of orthogonal states with finite amplitudes grows exponentially
in time assuming no interference between paths.  The spin-momentum basis for each neutrino
is very large and interference is essentially absent.  The amplitude of the initial state
is governed by unitarity and its amplitude decays quickly and hence the
one-body expectation values rapidly approach the equilibrium value.

We apply the algorithm to  given initial many-body product state. 
The interaction time for each pair can be sampled from
$\exp ( - \sum_j |V_{ij}| t)$, where the sum over j runs over all final
states with a finite matrix element to the initial state $i$.  After the interaction time
is determined for each pair, the chosen two-neutrino vertex
is determined by the shortest time.  After this vertex we have to choose
from all the various possible transitions for each pair.  We can choose these from the
normalized sum of the squares of all possible final states. In this manner a single
product state transitions to  a single different product state.

Each orthogonal final state remains orthogonal during further
time evolution due to the unitary nature of the time evolution.
Therefore, the unitary evolution is preserved once
we average over paths.  We have the full path history of each individual
flavor amplitude, allowing us to calculate any one-body observable
as a function of time, in particular, any possible correlations between
flavor and momentum or the number of neutrinos and antineutrinos of each 
flavor.  This is all we need since, in the astrophysical neutrino problem, one can really only measure one-body observables and the correlations between them.

We can then perform the diagonal propagation for the next two-neutrino vertex 
and repeat the entire procedure until we reach the desired final
time or distance. Note that in the next and subsequent steps, the interaction times of
particles not in the chosen pair are retained.  Those pairs involving one of the chosen
pair are resampled from the current time plus the sample of the inverse magnitudes of the updated state. The endpoint will be a product state of mass basis
neutrinos.

Of course, in general, there can be interference between paths, and if so, this semiclassical approximation will fail.  The fact that the neutrinos are non-degenerate with large kinetic energies 
means that correlations between different neutrino flavors and momentum rapidly decohere, leaving only interference between the flavor amplitudes of each individual neutrino.

\section{ Semiclassical versus full Hamiltonian simulations}

In Fig.~\ref{fig:neutrinodynamics}, we compare the evolution
of a small number of neutrinos (5 $\nu_e$, 3 $\nu_\mu$, and 2 $\nu_\tau$)
using $\exp [ - i H t] $  for the full full forward-scattering Hamiltonian 
to the semiclassical approximation for 100 neutrinos in the same ratio of flavors
and the same energy distribution for each flavor. 

We do not include the momentum degrees of freedom in the full Hamiltonian simulation.
The full quantum calculation is difficult with momentum 
because of the large number of states: 3 spin states times a very large number of momentum 
states for each neutrino.
The semiclassical calculation can be carried out for hundreds of
neutrinos at the same density. In this case we can also resample the
directions of the neutrinos at each vertex, here we sample the neutrino directions
randomly from a hemisphere each time they are scattered.

We choose an initial condition
so that the energy of the electron neutrinos varies uniformly from 4 to 6 MeV,
while the mu and tau neutrinos vary uniformly from 9 to 11 MeV.
The average initial electron neutrino energy is 5 MeV, while the $\nu_\mu$
and $\nu_\tau$ are 10 MeV.  
All neutrino directions are initialized randomly on a hemisphere.

For this initial calculation, we propagate for a total scaled time of 10 $\mu^{-1}$ with $\mu=\sqrt{2}G_F\rho_\nu$. The exchange matrix element is then
\begin{equation}
M_{ij} = \frac{G_F}{\sqrt{2}} \frac{\langle \lambda_i \cdot \lambda_j \rangle }{N}
\left(1-\hat{\bf k}_i\cdot\hat{\bf k}_j\right)\;.
\end{equation}

The time between non-diagonal interactions for an individual pair is distributed as
\begin{equation}
P_{ij}(t)   \propto  \exp\left(-\sum_{j\neq i}|M_{ij}| t\right)
\end{equation}
and is proportional to $1/N$ from the matrix element.  The next time for any transition involving particle i is then independent of N because there are N-1 pairs for each neutrino.

The shortest time among all pairs is used as the next interaction time. This time is
proportional to $1/N$ since there are N particles to choose from. 
Therefore even the semiclassical calculation is slightly more difficult for larger N. 
We can easily propagate of order 100 neutrinos on a laptop in this way.

In the limit of zero interference between different paths, the equilibration time
for one-body observables is independent of N (see Appendix~\ref{app:random_swaps}).  This result depends on having a vertex
that changes both flavor (spin) and momenta for a pair.
We have compared semi-classical calculations of 100 neutrinos to verify
this conclusion. Over a range of 10-100 neutrinos, 
the semiclassical results are independent of N within statistics.

Note that the energy in the flavor (spin) correlations is not conserved on its own,
but only when combined with the one-body Hamiltonian.  The energies in spins are of the order
of $10^{-10}$ that of $H_0$, so small variations
can easily be absorbed into tiny fluctuations of the neutrino momenta.
This is a significant difference from traditional mean-field
calculations on fixed grids particularly in 1D. It is similar to
fixed random couplings, but the evolution to the equilibrium state is faster.

The sampling procedure here does not conserve the magnitude of the total energy
dominated by the neutrinos kinetic energy,  unless one arbitrarily enforces energy
conservation at each vertex. Even this restriction would not drastically increase
the equilibration time, though. This is
similar to the fact that sampling exchanges do not conserve the total spin S.
The sampling of the pair momentum transfers could be changed to include kinetic energy
conservation, but realistically this should not be for each pair vertex; it should only apply to the entire system.
Total energy conservation
will not play a large role as long as we retain the correct frequency of vertices in the
path integral, since the magnitude of the momenta does not enter the spin (flavor) part of
the Hamiltonian.  Even single impurity evolution has a time scale as described above.
The evolution of the full system involves all pairs at each stage of the evolution, so it has
a similar time scale.  We note 
there are a very large number of nearly degenerate states with the same magnitude of
total kinetic energy for any arrangement of flavors (spins), with virtual excitations in the intermediate
states of the path integral. The exact correct frequency of vertices should be further explored, but in
any case, this time scale is much shorter than others in the problem.

\begin{figure}
\includegraphics[width=3.6in]{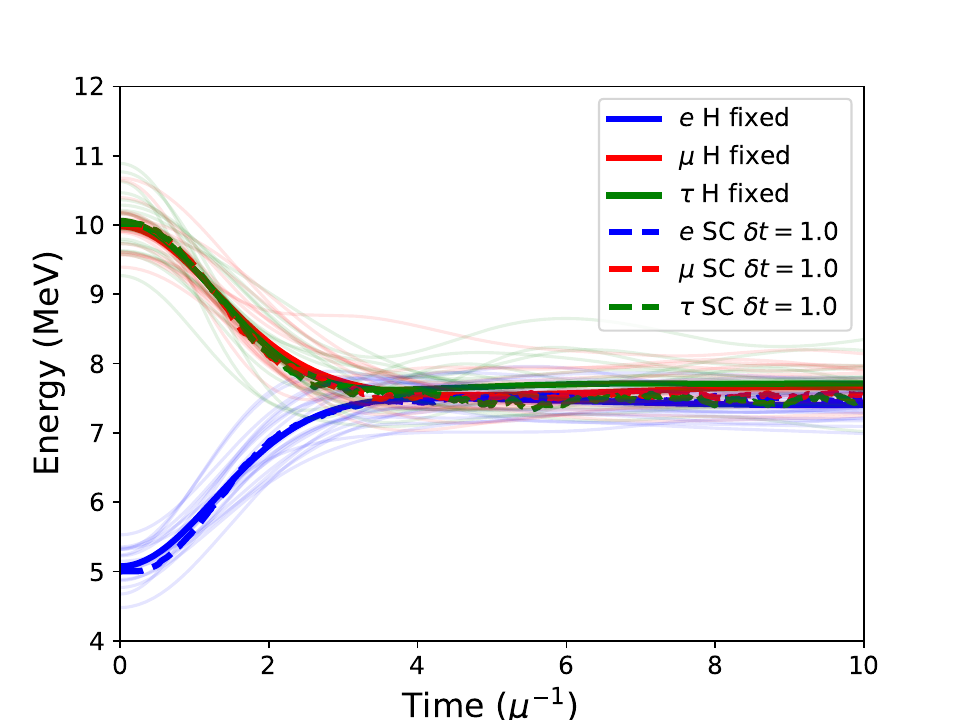}
\label{fig:neutrinodynamics}
	\caption{ Full semi-classical calculation for large N versus explicit many-body calculations of 10 neutrinos. Average of 10 many-body calculations are indicated by thick solid lines, with individual runs by fainter runs. Thick dashed lines are obtained by averages over many semiclassical runs with different samples of energies and momenta for 100 neutrinos. The full H calculation is an average over instances of fixed couplings, while the semiclassical runs resample the couplings at a rate of 1 $\mu^{-1}$. }
\end{figure}

In Fig.~\ref{fig:neutrinodynamics} we compare the semiclassical
calculations to full many-body quantum simulations with random couplings on a hemisphere.
We use only 10 neutrinos for the full calculations, 
as each can be in one of 3 flavor states, yielding a large number of total amplitudes.  The initial couplings are chosen from a uniform neutrino directions on the hemisphere.  These directions and the couplings are kept fixed in this calculation,
since it is not possible to match a spin Hamiltonian to the full Hamiltonian exactly.
The semiclassical method can randomly resample couplings  at each interaction vertex.
For $N=10$, the oscillations at larger times are due to the small number of neutrinos, and are at a very small scale compared to the
original energy differences. The thick solid lines are obtained as an average over 10 full Hamiltonian
calculations with different samples for the initial condition.  The faint lines indicate individual many-body calculations.

The thick dashed lines are averages over 100 semiclassical calculations with 100 neutrinos each.  In these calculations
we sample the first N/2 time steps from each run from the greater of two samples of the magnitude of the matrix element, which results in a 
quadratic behavior near T=0. The initial alignment of flavor states
requires both left- and right-hand paths in the path integral to flip to
change the expectation value; thus, so picking the larger of two times for the chosen pair gives
the correct quadratic behavior in time for the correlation function. After the first $\sim N/2$ exchanges
, the spins are no longer in flavor eigenstates; therefore, so we just sample from a single exponential.
To gauge the effect of the varying couplings, the updated directions of the momenta are chosen randomly from the hemisphere (the same sampling as
for the initial configuration). 

The semiclassical calculations give the correct equilibrium distribution,
the correct time scale for the equilibration, and can easily accommodate
 hundreds up to  one thousand neutrinos.  These results should be contrasted with previous full many-body calculations performed considering only $2$ neutrino flavors and only the forward-scattering interaction: these lead to typical evolution time-scales growing with system size depending on the initial condition and the geometry of the system~\cite{Roggero2021a,Roggero2021b,Roggero2022,Martin:2021bri}. 

The equilibration time for one-body observables can be different than complete 
equilibration.  In addition,
different  time scales  for the two-body interactions and the
change of coupling may give different scaling results. For the spin-only Hamiltonian,
very slow changes will lead to
the growth of many-body equilibration time with N since the energy variance of the initial
product state is proportional to $\sqrt{N}$ for our Hamiltonian, where the total
spectral width is proportional to N. General equilibration cannot be faster than the
inverse of this variance.

A slow change in the spin Hamiltonian will essentially reproduce fixed Hamiltonian results.
Very rapid changes will give a diffusion of the spin coordinates, again slowing
equilibration. In the physical system considered here, the coupling changes
at each vertex. Some two-neutrino vertices do not change flavor, but they are 
comparable to or smaller than those that change both momentum and flavor, 
resulting in changes in coupling on time scales comparable to the spin term.
Extremely fast changes will induce a diffusion like scaling for equilibration.

We find that one-body observables at constant density equilibrate within a constant time as a function of N
for the full Hamiltonian with its implicit time dependence. For a time-varying Hamiltonian, the full many-body equilibration time is expected to be logarithmic in N, while for fixed Hamiltonians it should increase as a power of N.  The randomness reduces the size dependence of the one-body equilibration.

There is a strong similarity between this problem and the problem of random quantum circuits
, much discussed in condensed matter~\cite{Fisher-2023}.
These circuits are often very fast scramblers~\cite{Brown-2012}, particularly for the
one-body observables we consider here.  In our case, there are important symmetries that
need to be included (conservation of neutrino and antineutrino number, number of neutrinos minus antineutrinos of each flavor, etc.)  The very rapid equilibration arises because the
neutrinos are non-degenerate; hence, there is no Pauli blocking.  The short-range interaction allows coupling to many possible transitions at each time, and the kinetic energy of the neutrino momentum dominates energy conservation, enabling large fluctuations in the flavor
energy.
 
As shown in Ref.~\cite{Martin:2023}
, the neutrino-antineutrino interactions also lead to rapid equilibration
in the neutrino and antineutrino densities of each flavor.  
As we discuss below, this equilibration is easiest to see with very
large matter background densities or for nearly vacuum conditions.
In these cases, the flavor eigenstates of the neutrinos and antineutrinos
are identical.  Below, we consider modest constant background densities
and varying densities governed by MSW evolution, where the background
matter density is changing.

\section{Modest Matter Densities}
\label{sec:mod_matter_dens}

Unlike the vacuum case or the case of large matter densities, at more modest densities
, the mass eigenstates of neutrinos and antineutrinos are not identical.
In such a case, we cannot rotate the Hamiltonian to a basis where the neutrino-antineutrino
interaction simply changes the flavor of the final pair relative to the initial pair.
One can only rotate to either the neutrino mass eigenstates for both neutrinos and antineutrinos,
or to the antineutrino mass eigenstates for both.

It is simplest to consider rotating both to the majority eigenstates, either neutrinos for
the $N_\nu>N_{\bar{\nu}}\,$
or antineutrino mass eigenstates for $N_\nu<N_{\bar{\nu}}$ instead.
The neutrino-antineutrino interaction will then cause a very rapid equilibration in this basis,
but the minority species is not in a one-body mass basis and will slowly evolve.
It may be that the background densities are evolving more rapidly than this,
and hence one should consider the MSW case with spatially varying densities.

Full equilibration will occur on scales associated with the mass differences squared of the
minority species. If only one background density is large and only one mass state is identical for neutrinos and antineutrinos, it should be possible to define an equilibrium between the product of densities of this flavor and the sum of the products of the other
flavors, where the neutrino and antineutrino states do not coincide.

\subsection{MSW conversions including neutrino-neutrino interactions}

The results of Ref.~{\cite{Martin:2023}}
demonstrate the rapid equilibration of momenta (magnitude and angles) of different flavors of neutrinos and antineutrinos in the presence of a sizable background matter potential or in the vacuum.  The MSW matter potential yields neutrino eigenstates that correspond to flavor eigenstates for electron neutrinos and antineutrinos and for all flavors if there is a significant net muon contribution to the background. It is 
important to understand what happens as the matter background decreases at larger distances. Note that non-forward scattering may result in angular distributions that do not
reduce to the pure forward scattering limit of $1/R^4$. In this case, two-neutrino
interactions may be more important than those for forward scattering.

We first consider the simplest case of neutrinos only going through an MSW resonance. Without neutrino-neutrino interactions, the standard MSW evolution occurs, with potential dependence on the neutrino energy as the resonance occurs at slightly different densities for different energies.  With neutrino-neutrino interactions, one can track the evolution of each flavor of neutrino, with momenta being exchanged through the
two-neutrino interactions.  The evolution of each initial energy neutrino then becomes independent of energy, and the neutrino evolution for each neutrino is equal to the standard MSW evolution with an inverse energy equal to the average energy of the neutrinos. Note that the exchanges conserve this average energy over all neutrinos. We have verified this result with explicit semiclassical calculations.

We assume a specific initial condition where the electron neutrinos dominate as expected, for example, during the neutronization burst (see ~\cite{Duan-Nburst2008} and references therein).  Without oscillations, the energies of the electron neutrinos are expected to be, on average, lower than those of the other flavors, as they decouple later.
However, the neutrino neutrino interactions rapidly equilibrate the neutrino and antineutrino momenta.
We take as an example a state that is already equilibrated by passing through a high-density region.
We take neutrinos and anti-neutrinos in initial conditions with flux ratios given by 9:1:3:3:3:3 for $\nu_e$:$\bar{\nu}_e$:$ \nu_{\mu}$:$\bar{\nu}_{\mu}$:$\nu_{\tau}$:$ \bar{\nu}_{\tau}$.
We consider two limiting cases for the evolution of the matter density: a rapid quench in the matter potential and a slow adiabatic reduction of the background density.  These two limiting cases provide a picture of what will happen in general.

The rapid quench case is simple in that we can consider the instantaneous overlap of the mass eigenstates before and after the quench.  Here we take a quench to zero matter density, so we use vacuum eigenstates after the quench.  The initial distributions will be governed by the equilibration described above: equal densities for the product of neutrino and antineutrino mass eigenstates (at high densities, also flavor eigenstates) and equal energy distributions.
For an instantaneous quench, we can simply calculate
the instantaneous transition to the vacuum eigenstates.  These new distributions in terms of the vacuum eigenstates will then rapidly equilibrate through the change of flavors ( mass eigenstates) of neutrino/antineutrino pairs.

\begin{figure}
\includegraphics[width=3.75in]{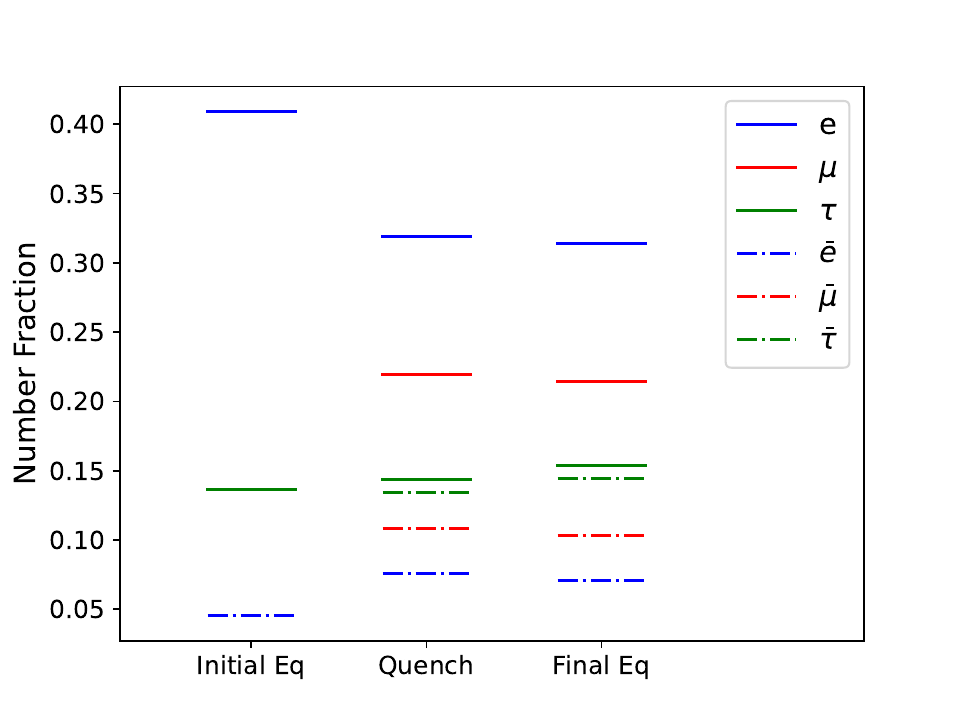}
\label{fig:neutrinodynamics}
\caption{ Dynamics for the rapid quench of the matter density. The initial distribution is shown on the left, this corresponds to the vacuum flavor distribution shown in the middle, which equilibrates to the relative densities shown on the right.}
\label{fig:quench}
\end{figure}

The quench result for the initial conditions described above is shown in Fig. \ref{fig:quench}. 
The initial densities are on the left, corresponding to vacuum densities shown in the middle, and final equilibrated densities on the right. The difference between electron neutrino and antineutrino densities become smaller, while the others are largely unchanged.  This is due to the initial and final overlaps of the mass eigenstates at high density and zero density.  For these calculations, we assume a normal mass hierarchy.

At the other extreme, we consider slow evolution of the matter densities for the same initial conditions.
Here we show results for an exponentially decaying
profile of the form $\rho_e- \rho_{\bar{e}} =
\rho_0  \exp [ - (r/s)]$,  starting at 150 km and with a scale height of 25 km. and neutrinos and antineutrinos equilibrating to an average energy of 5 MeV. 

\begin{figure}
\includegraphics[width=3.6in]{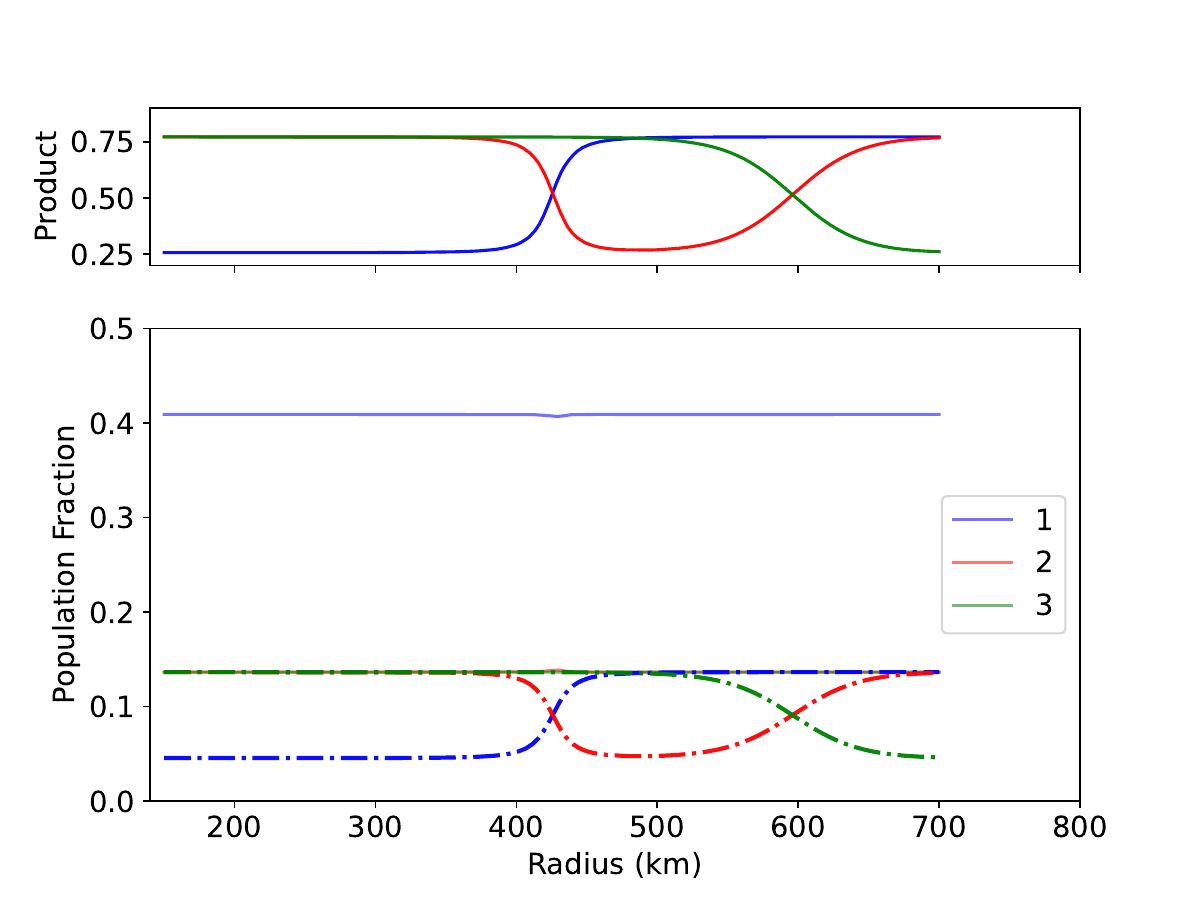}
\caption{ Standard MSW in the neutrino mass basis for both neutrinos and antineutrinos. Neutrinos are shown as solid lines, antineutrinos as dash-dot lines. The neutrino evolution is nearly adiabatic (constant), the antineutrino density is nearly adiabatic in the antineutrino basis, but evolves with time in the neutrino mass basis, breaking the flavor equilibration arising from neutrino-antineutrino interactions.} \label{fig:allmassbasis}
\end{figure}

The standard MSW evolution in this non-standard anti-neutrino basis is shown in Fig. \ref{fig:allmassbasis}.
The lower panel shows neutrino (solid lines) and antineutrino densities (dashed lines) in the
neutrino mass basis. Neutrino densities are nearly flat  as the evolution
is close to adiabatic.  The antineutrino densities would also be nearly flat if plotted in the antineutrino
mass basis.  Here, though, we transform them to the neutrino mass basis to see how the initial equilibrated
state evolves under MSW evolution alone..
The upper panel shows the product of neutrinos and antineutrinos in the 
neutrino mass basis.  This shows a significant deviation from equilibration once the neutrino and antineutrino
mass eigenstates do not align, for this case around 300 km. 

The initial state has a relatively large number of electron neutrinos and a small number of electron antineutrinos, represented by solid and dashed blue lines in the figure.
For neutrinos, this is the highest mass state at high density; since the initial number of electron antineutrinos is small
, the corresponding density (blue dashed line) is also small. 
As expected in an adiabatic conversion, the highest mass state remains the highest mass state as we evolve to the final vacuum Hamiltonian, their flavor changes from electron to mu or tau.  The mu and tau neutrinos are converted to all flavors. 
The antineutrinos do not undergo an MSW resonance in this case; however, the fact that the neutrino states do implies that the
antineutrinos evolve when plotted in the neutrino mass basis. 
The antineutrino flavor densities are evolving, as the densities of antineutrinos in the neutrino mass basis are not constant. At large $r$, starting around 300 km, the deficit changes from electron flavor
to $\mu$ flavor and then to $\tau$ flavor, corresponding to neutrino basis MSW resonances in electrons to the other flavors. 

\begin{figure}
\includegraphics[width=3.6in]{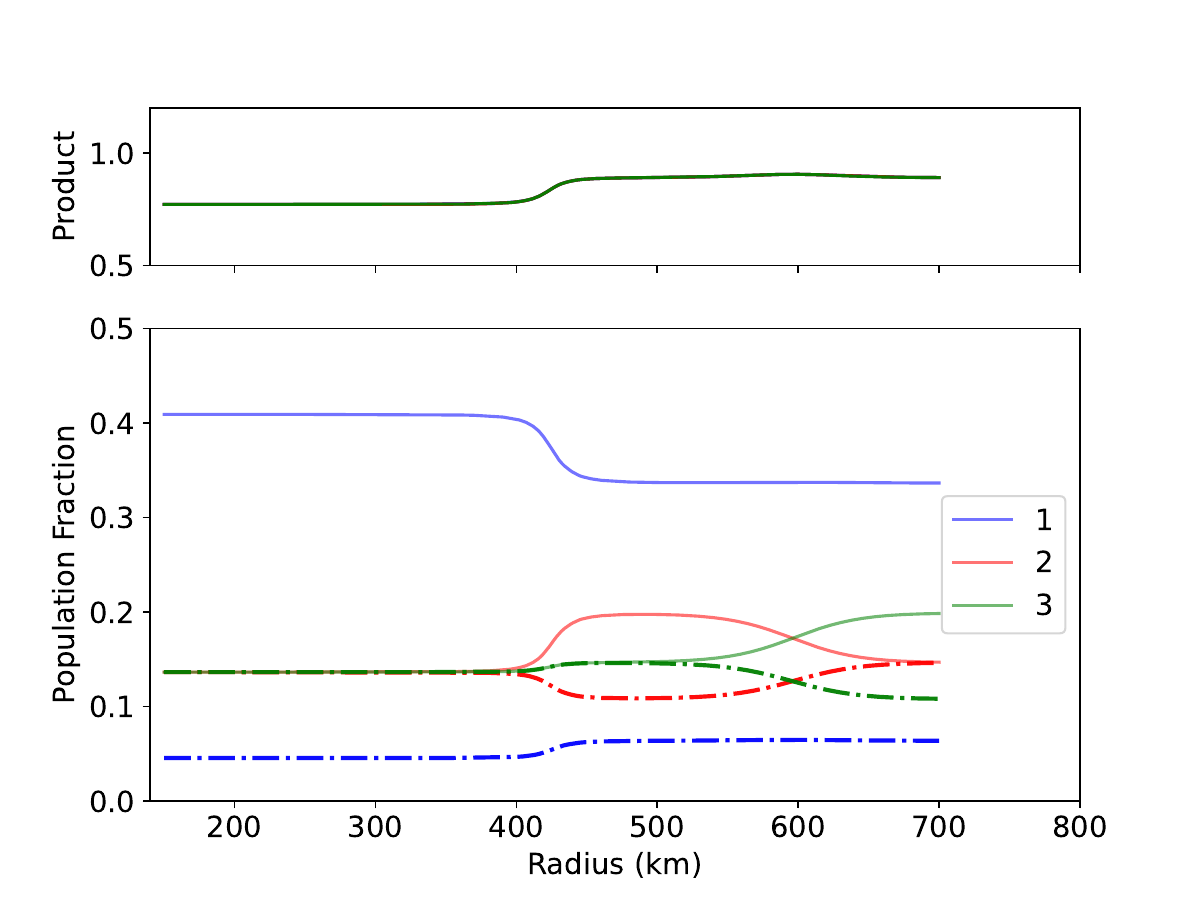}
\caption{ Standard MSW plus flavor equilibration
in the neutrino mass basis for both neutrinos and antineutrinos. Both neutrino and antineutrino densities evolve with time due to the flavor equilibration.} 
\label{fig:eqallmassbasis}
\end{figure}

If the time scale of the neutrino-antineutrino coupling is 
very small compared to density variations in MSW, we can
calculate its effect by equilibrating the product of densities
in the instantaneous neutrino mass basis. The results of this equilibration
are shown in the neutrino mass basis in Fig. \ref{fig:eqallmassbasis}.  
Since these densities are initially equilibrated, the rapid transitions of the ${\nu}_e$
density are offset to some degree by flux coming back from
$\nu_{\mu}$, $\bar{\nu}_{\mu}$, and $\nu_{\tau}$, $\bar{\nu}_{\tau}$.

\begin{figure}
\includegraphics[width=3.5in]{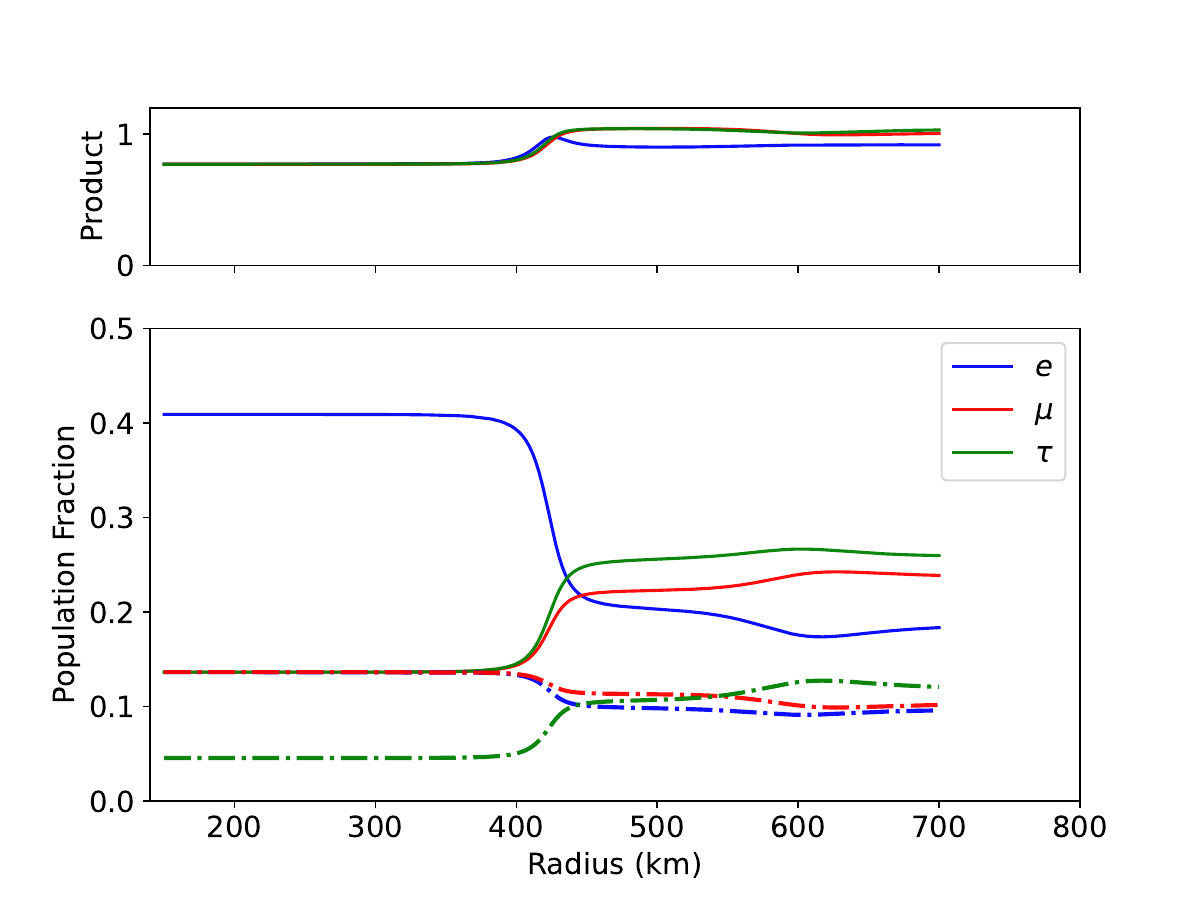}
\caption{ Standard MSW plus equilibration in the  flavor basis. The upper panel shows
a slight remaining imbalance because we are equilibrating in the neutrino mass basis for the antineutrinos (fast equilibration)
and not including the slower equilibration from pure flavor equilibration (see text).} 
\label{fig:eqflavorbasis}
\end{figure}

Transforming these results directly to the flavor basis for neutrinos and antineutrinos is shown in Fig. \ref{fig:eqflavorbasis}.
The electron neutrino density is reduced, but not as much as in the MSW only calculation. A small difference in the product of flavor densities remains because we have only included the fast equilibration in the neutrino flavor basis, not the slower equilibration
related to the single particle antineutrino evolution in its own mass basis rather than the neutrino mass basis (see text).

\begin{figure}
\includegraphics[width=3.5in]{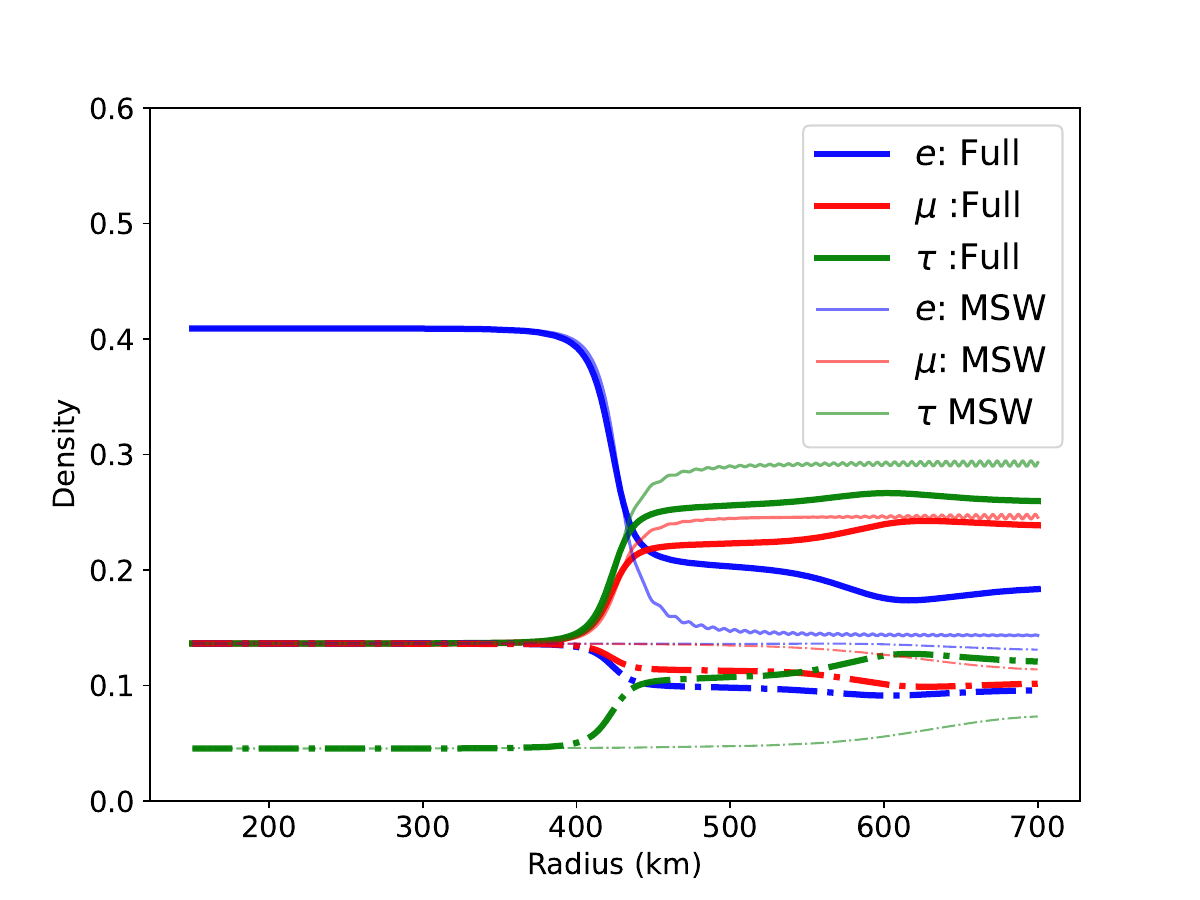}
\caption{ Comparison of MSW plus neutrino-antineutrino evolution to MSW alone. } 
\label{fig:flavorMSWeq}
\end{figure}

Finally, we compare the flavor densities for neutrinos and antineutrinos with MSW plus neutrino-antineutrino 
equilibration to the MSW evolution  alone in Fig. \ref{fig:flavorMSWeq}. 
In the figure, solid lines represent neutrinos, while dashed lines represent antineutrinos.  MSW only results are shown as thin lines, while
MSW plus equilibration are represented as thicker lines. The flavor transition probabilities of the majority (electron neutrinos) are reduced
by flavor equilibration.  The MSW evolution reduces the electron neutrino density, but this reduction is partially
compensated by neutrino-antineutrino interactions. The minority anti-neutrinos change fractionally more than the neutrinos since each vertex
converts one pair to another flavor or mass state.  In the limit of large ratios of neutrinos to antineutrinos, the evolution
of the majority (neutrinos or antineutrinos) reduces to the standard MSW result.

The final results are closer to full equilibration among all flavors than MSW alone, but the impact
is reduced for large or small neutrino to antineutrino ratios.
The main impact is to somewhat inhibit MSW conversions, this impact
will be larger for more equal neutrino and antineutrino populations. 

\section{Conclusions}
The essential features of neutrino evolution in astrophysical environments can be described in a 
semiclassical approach due to the fast decoherence time scales. These
time scales would only be quicker by adding production or absorption
processes due to their incoherent nature.

Decoherence leads to very fast equilibration between the energies of different
flavors, essentially independent of the initial angular distributions of each
flavor. In regions with high mass density, the number of neutrinos
minus the number of antineutrinos in each mass eigenstate is conserved. 
The neutrino-antineutrino
flavor-changing processes lead to an equilibration in the product of
the neutrino and antineutrino densities of the different species.
For this reason, a significant memory of the initial state will be retained,
particularly when there is initially a large imbalance between the total numbers of neutrinos and antineutrinos.

The impact of this equilibration in neutron star mergers has been studied in Ref.~\cite{qiu:2025}.
They find that  neutrino flavor transformations impact the composition and structure of the remnant, potentially leaving an imprint on the postmerger gravitational-wave signal. They also have a significant impact on the composition and nucleosynthesis yields of the ejecta.

The neutronization burst in a supernova will produce many more electron neutrinos than antineutrinos~. In contrast, in a neutron star merger, many neutrons
are converted into protons, leading to a large production of electron
anti-neutrinos.  In each case, a strong memory of the initial conditions
will be retained, and the energy of the electron neutrinos and antineutrinos will equilibrate with the other species, leading to a potentially significant
effects on energy deposition in matter and nucleosynthesis.

Fast flavor instabilities in core collapse supernovae have been studied in several works, including
\citenum{wang2025effectfastflavor}, \citenum{wang2025effectcollisional}. and \citenum{wang2025instabilitycorecollapse}.
These have often been performed incorporating mean-field approaches for the neutrino dynamics, which
evolve quickly but are quite distinct from the evolution we obtain. In particular they do our results
do not depend on crossing in angular distributions, which in any case here equilibrate very quickly with
collisions.  In addition, the final flavor distributions are different.  It is not clear how these differences impact different regions of the supernovae.  These studies have demonstrated that nucleosynthesis can be impacted.

This memory will be partially erased further out from the neutrinosphere during adiabatic or near adiabatic
MSW transitions, leading to a  more constant ratio of neutrinos to antineutrinos
in all flavors. In the case of core-collapse supernovae, this prediction could in principle be checked by terrestrial observations and/or nucleosynthesis observations.

This approach leads to many important topics not considered here, including
the impact (if any) of Majorana neutrinos versus the Dirac neutrinos
considered here.  More studies on the impact of the hierarchy are warranted,
as well as various types of initial conditions. A quantitative characterization of the equilibration timescale in regions with a large neutrino flux will also require further studies, particularly in assessing the role played by many-body correlations, which are directly responsible for neutrino thermalization. Finally, studies of the
impact of neutrinos in the early universe could be revisited.

\section{Acknowledgements}

We thank Vincenzo Cirigliano and George
Fuller for valuable conversations. This work was supported by the Quantum Science Center (QSC), a National
Quantum Information Science Research Center of the U.S.
Department of Energy (DOE), and by the U.S. Department of
Energy, Office of Science, Office of Nuclear Physics (NP)
Contract No. DE-AC52-06NA25396. 

\appendix

\section{Two Neutrino Phase Space and Energy Conservation}
\label{app:twob_vertex}

  In this appendix, we describe  
the two-body phase space, including energy conservation
for final states of two neutrinos
with 3-momenta ${\bf p}_1$ and ${\bf p}_2$, conserving momenta and energy.
We assume the matrix elements are short-range and thus independent of the magnitude of the vector three-momentum transfer ${\bf q}$.
The total momentum is  ${\bf P} = {\bf p}_1 + {\bf p}_2$;
we can also calculate the magnitudes of the parallel and perpendicular
components of each momentum relative to ${\bf P}/2$:
\begin{eqnarray}
p_{||}  & = &  {\bf p}_i \cdot {\hat {\bf P}}  - {\bf P}/2 \nonumber \\
p_{\perp} & = &  {\bf p}_i \times  {\hat {\bf P}} 
\end{eqnarray}
Without loss of generality, we can rotate so that ${\bf P}$ is aligned with the
$z$-axis and ${\bf p}_{\perp}$ is aligned with the $x$-axis.  The two initial lab frame momenta are:
\begin{eqnarray}
	{\bf p}_1  & = & {\bf P}/2 + {\bf p}_{||} + {\bf p}_{\perp} \nonumber \\
	{\bf p}_2  & = & {\bf P}/2 - {\bf p}_{||} - {\bf p}_{\perp}
\end{eqnarray}
For the final state of the two-neutrino system, we need to enforce momentum conservation and energy conservation.  The energy of the initial state is:
\begin{equation}
\begin{split}
E_{init}  =& | {\bf p}_1 | + | {\bf p}_2|  \\
  = &\sqrt { ({\bf P}/2 + p_{||})^2 + p_{\perp}^2}\\
&+ 
	\sqrt { ({\bf P}/2 - p_{||})^2 + p_{\perp}^2}
\end{split}
\end{equation}
The energy of the final state is written similarly in terms of final state  parallel and perpendicular momenta ${\bf p'}_{||}$ and ${\bf p'}_{\perp}$.
Since the contribution of ${\bf p'}_\perp$ to the energy is positive,
the maximum magnitude of  ${\bf p}_{||}$  is found by solving $E_{fin} = E_{init}$ for ${\bf p'}_{\perp} = 0$. Similarly, the maximum magnitude of ${\bf p}_{\perp}$is obtained when ${\bf p'}_{\perp}$ = 0.  These equations yield:
\begin{eqnarray}
	p'_{max,||} & = &  E_{init}/2 \nonumber \\
	p'_{max,\perp} & = & \sqrt{(E_{init}^2 - 2 ({\bf P}/2)^2}
\end{eqnarray}

We show a specific example of the impact of the energy conserving delta function in Figure \ref{fig:vertices-mom}. The momenta of the two initial particles are shown as the two large dots, and the lines represent all possible final states that conserve two-neutrino energy and momenta.

If we call these maximum
magnitudes $p_{max,||}$ and $p_{max,\perp}$, respectively, we can compute
the phase space integral as
\begin{eqnarray}
PS & = & \int d^3 \mathbf{q}\, \delta ( E_{final} - E_{init}) \nonumber \\ 
	& = &  (2 \pi) \  0.785  \ p_{max,||} \  p_{max,\perp},
\label{eq:finalstates}
\end{eqnarray}
where the $2 \pi$ comes from the azimuthal symmetry, the two maximum momenta can be calculated from the initial momenta, and the 0.785 comes from a numerical integration of the energy conserving delta function. The units are energy, or momentum, squared, as expected for massless particles. We have used a short-range interaction where the matrix element is independent of the magnitude of the momentum transfer squared.

In the semiclassical approximation, we enforce energy conservation after
each neutrino-neutrino vertex. 

The total rate for two neutrinos can be obtained by summing over final states of the
pair imposing energy conservation:
\begin{equation}
\begin{split}
{\rm Rate} &=\frac{2\pi}{\hbar c}\sum_{f,x,q} | \langle H_{fi} \rangle |^2  \delta(E_f - E_i)\\
&=\frac{2\pi}{\hbar c} \frac{V}{(2\pi\hbar c)^3}\int d^3\mathbf{q} \,\langle H_{fi} \rangle |^2  \delta(E_f - E_i)\\
&=\frac{V}{4\pi^2(\hbar c)^4}\int d^3\mathbf{q}\left[\sum_{f,x}| \langle H_{fi} \rangle |^2 \right] \delta(E_f - E_i)\\
\end{split}
\end{equation}
but then we approximate
\begin{equation}
\sum_{f,x}| \langle H_{fi} \rangle |^2\approx \left[(\hbar c)^3\frac{\sqrt{2}G_F}{V}\right]^2\;,
\end{equation}
and then the final expression for the rate looks like
\begin{equation}
\begin{split}
{\rm Rate}&=\frac{V}{4\pi^2(\hbar c)^4}\frac{(\hbar c)^62G_F^2}{V^2}\int d^3\mathbf{q} \delta(E_f - E_i)\\
&=\frac{(\hbar c)^2G^2_F}{2\pi^2}\frac{\rho}{N}PS
\end{split}
\end{equation}

For two neutrinos the cross section can be obtained from the matrix element squared
by including both the energy conserving delta function and the relativistic "Moller"
flux factor~\cite{nachtmannelementary,hagedorn1962selected}.

\section{Random SWAPS}
\label{app:random_swaps}

Here we want to show that if we take a neutrino system and we apply random SWAP operations to them, the system will reach equilibrium in the one-body observables exponentially fast with a rate which is independent of the size $N$ of the system.

The evolution is as follows: at every time-step we pick a pair of neutrinos $(i,j)$ at random and apply the operation $P_{ij}$ to the state; this process is then repeated for $L$ layers. The idea is to perform the calculation in the Heisenberg picture and evolve the observable instead of the state. We define equilibration once the individual flavor observables $\langle Z_i(t)\rangle$ are all equal to their average $Z_{av}=\frac{1}{N}\sum_i\langle Z_i(t)\rangle=\frac{1}{N}\sum_i\langle Z_i(0)\rangle$ where the equality follows from the conservation of the total $Z$ in the system. Since all the expectation values go to the same limit, we consider the evolution of $Z_1$ only. We use $U_k$ to specify the unitary applied at layer $k$. After one one layer we have
\begin{equation}
Tr\left[\rho_0 U^\dagger_1Z_1U_1\dagger\right] = (1-p) \langle Z_1(0)\rangle+p\langle Z_{1_1}(0)\rangle\;,
\end{equation}
where $\rho_0$ is the initial state, $1_1$ is the label of the neutrino swapped with $1$ at that the first neutrino after one layer and $p$ is the probability to select neutrino $1$ at this layer. For uniform sampling we get $p=(N-1)/\binom{N}{2}=2/N$. If we do an average over realizations we then get
\begin{equation}
\mathbb{E}Tr\left[\rho_0 U^\dagger_1Z_1U_1\right] = (1-p) \langle Z_1(0)\rangle+p Z_{av}\;.
\end{equation}

After two layers, the probability of staying in $1$ is given by the probability $(1-p)^2$ of staying there for both layers plus the probability $p/\binom{N}{2}$ of jumping once and then back to $1$. We then have
\begin{equation}
\begin{split}
\mathbb{E}Tr\left[\rho_0 U^\dagger_1U^\dagger_2Z_1U_2U_1\right] &= \left[(1-p)^2+\frac{p}{\binom{N}{2}}\right] \langle Z_1(0)\rangle\\
&+\left[p^2+2p(1-p)-\frac{p}{\binom{N}{2}} \right]Z_{av}\;.
\end{split}
\end{equation}

For a large system when $\binom{N}{2}\gg 1/(1-p)\gg1/p$ we can approximate this neglecting factors of $p/\binom{N}{2}$ (at least if the total number of layers $L\ll\binom{N}{2}$). But then, after $L$ layers we have
\begin{equation}
\begin{split}
\mathbb{E}Tr\left[\rho_0 (\prod_{k=1}^LU^\dagger_k)Z_1(\prod_{k=L}^1U_k)\right] &= (1-p)^L \langle Z_1(0)\rangle\\
&+\left[1-(1-p)^L \right]Z_{av}\;.
\end{split}   
\end{equation}
But then this means that
\begin{equation}
\begin{split}
\Delta_L=&\mathbb{E}Tr\left[\rho_0 (\prod_{k=1}^LU^\dagger_k)Z_1(\prod_{k=L}^1U_k)\right]-Z_{av}\\
=&(1-p)^L \left[\langle Z_1(0)\rangle - Z_{av}\right]\;,
\end{split}
\end{equation}
which we can bound as
\begin{equation}
|\Delta_L|<2(1-p)^L=2\left(1-\frac{2}{N}\right)^{L}
\end{equation}
But then, the number of layers required to converge to a flavor expectation value, on each individual neutrino, which is on average at most $\delta$ away from the average $Z_{av}$ is at most
\begin{equation}
L_{eq}=-\frac{\log(1/\delta)}{\log(1-p)}\gtrsim p\log\left(\frac{2}{\delta}\right)=\frac{N}{2}\log\left(\frac{2}{\delta}\right)\;.
\end{equation}

To connect this with a time scale, we assume $H=(\mu/N)\sum_{i<j}c_{ij}P_{ij}$ with coupling constants $c_{ij}\geq0$ sampled from some distribution. The incremental time step $\Delta \tau$ at every layer is sampled from the shortest collision time on all the pairs. It's average value over different runs is then
\begin{equation}
\mathbb{E}[\Delta \tau]=\frac{N}{\mu\sum_{i<j}c_{ij}}\;,
\end{equation}
and if we average this over the choices of coupling constants (ie. sampling one random set for each run and keeping them for the $L$ layers) gives
\begin{equation}
\mathbb{E}_C\mathbb{E}[\Delta \tau]=\mathbb{E}_C\left[\frac{N}{\mu\sum_{i<j}c_{ij}}\right]\geq \frac{N}{\mu\mathbb{E}_C[\sum_{i<j}c_{ij}]}\;.
\end{equation}
If we choose directions uniformly than we showed some time ago that $\mathbb{E}_C[\sum_{i<j}c_{ij}]=\binom{N}{2}$ so that the expected equilibration time goes as
\begin{equation}
T_{eq}=L_{eq}\mathbb{E}_C\mathbb{E}[\Delta \tau]\gtrsim \frac{1}{\mu}\frac{N}{N-1}\log\left(\frac{2}{\delta}\right)\;.
\end{equation}
Of course this is completely independent on what the initial state was.


\begin{thebibliography}{70}%
\makeatletter
\providecommand \@ifxundefined [1]{%
 \@ifx{#1\undefined}
}%
\providecommand \@ifnum [1]{%
 \ifnum #1\expandafter \@firstoftwo
 \else \expandafter \@secondoftwo
 \fi
}%
\providecommand \@ifx [1]{%
 \ifx #1\expandafter \@firstoftwo
 \else \expandafter \@secondoftwo
 \fi
}%
\providecommand \natexlab [1]{#1}%
\providecommand \enquote  [1]{``#1''}%
\providecommand \bibnamefont  [1]{#1}%
\providecommand \bibfnamefont [1]{#1}%
\providecommand \citenamefont [1]{#1}%
\providecommand \href@noop [0]{\@secondoftwo}%
\providecommand \href [0]{\begingroup \@sanitize@url \@href}%
\providecommand \@href[1]{\@@startlink{#1}\@@href}%
\providecommand \@@href[1]{\endgroup#1\@@endlink}%
\providecommand \@sanitize@url [0]{\catcode `\\12\catcode `\$12\catcode `\&12\catcode `\#12\catcode `\^12\catcode `\_12\catcode `\%12\relax}%
\providecommand \@@startlink[1]{}%
\providecommand \@@endlink[0]{}%
\providecommand \url  [0]{\begingroup\@sanitize@url \@url }%
\providecommand \@url [1]{\endgroup\@href {#1}{\urlprefix }}%
\providecommand \urlprefix  [0]{URL }%
\providecommand \Eprint [0]{\href }%
\providecommand \doibase [0]{http://dx.doi.org/}%
\providecommand \selectlanguage [0]{\@gobble}%
\providecommand \bibinfo  [0]{\@secondoftwo}%
\providecommand \bibfield  [0]{\@secondoftwo}%
\providecommand \translation [1]{[#1]}%
\providecommand \BibitemOpen [0]{}%
\providecommand \bibitemStop [0]{}%
\providecommand \bibitemNoStop [0]{.\EOS\space}%
\providecommand \EOS [0]{\spacefactor3000\relax}%
\providecommand \BibitemShut  [1]{\csname bibitem#1\endcsname}%
\let\auto@bib@innerbib\@empty
\bibitem [{\citenamefont {Wolfenstein}(1978)}]{PhysRevD.17.2369}%
  \BibitemOpen
  \bibfield  {author} {\bibinfo {author} {\bibfnamefont {L.}~\bibnamefont {Wolfenstein}},\ }\href {\doibase 10.1103/PhysRevD.17.2369} {\bibfield  {journal} {\bibinfo  {journal} {Phys. Rev. D}\ }\textbf {\bibinfo {volume} {17}},\ \bibinfo {pages} {2369} (\bibinfo {year} {1978})}\BibitemShut {NoStop}%
\bibitem [{\citenamefont {Mikheyev}\ and\ \citenamefont {Smirnov}(1989)}]{MIKHEYEV198941}%
  \BibitemOpen
  \bibfield  {author} {\bibinfo {author} {\bibfnamefont {S.}~\bibnamefont {Mikheyev}}\ and\ \bibinfo {author} {\bibfnamefont {A.}~\bibnamefont {Smirnov}},\ }\href {\doibase https://doi.org/10.1016/0146-6410(89)90008-2} {\bibfield  {journal} {\bibinfo  {journal} {Progress in Particle and Nuclear Physics}\ }\textbf {\bibinfo {volume} {23}},\ \bibinfo {pages} {41} (\bibinfo {year} {1989})}\BibitemShut {NoStop}%
\bibitem [{\citenamefont {Pantaleone}(1992{\natexlab{a}})}]{pantaleone1992neutrino}%
  \BibitemOpen
  \bibfield  {author} {\bibinfo {author} {\bibfnamefont {J.}~\bibnamefont {Pantaleone}},\ }\href@noop {} {\bibfield  {journal} {\bibinfo  {journal} {Physics Letters B}\ }\textbf {\bibinfo {volume} {287}},\ \bibinfo {pages} {128} (\bibinfo {year} {1992}{\natexlab{a}})}\BibitemShut {NoStop}%
\bibitem [{\citenamefont {Pantaleone}(1992{\natexlab{b}})}]{PhysRevD.46.510}%
  \BibitemOpen
  \bibfield  {author} {\bibinfo {author} {\bibfnamefont {J.}~\bibnamefont {Pantaleone}},\ }\href {\doibase 10.1103/PhysRevD.46.510} {\bibfield  {journal} {\bibinfo  {journal} {Phys. Rev. D}\ }\textbf {\bibinfo {volume} {46}},\ \bibinfo {pages} {510} (\bibinfo {year} {1992}{\natexlab{b}})}\BibitemShut {NoStop}%
\bibitem [{\citenamefont {Qian}\ and\ \citenamefont {Fuller}(1995)}]{Qian:1994wh}%
  \BibitemOpen
  \bibfield  {author} {\bibinfo {author} {\bibfnamefont {Y.~Z.}\ \bibnamefont {Qian}}\ and\ \bibinfo {author} {\bibfnamefont {G.~M.}\ \bibnamefont {Fuller}},\ }\href {\doibase 10.1103/PhysRevD.51.1479} {\bibfield  {journal} {\bibinfo  {journal} {Phys. Rev.}\ }\textbf {\bibinfo {volume} {D51}},\ \bibinfo {pages} {1479} (\bibinfo {year} {1995})},\ \Eprint {http://arxiv.org/abs/astro-ph/9406073} {arXiv:astro-ph/9406073 [astro-ph]} \BibitemShut {NoStop}%
\bibitem [{\citenamefont {Pastor}\ \emph {et~al.}(2002)\citenamefont {Pastor}, \citenamefont {Raffelt},\ and\ \citenamefont {Semikoz}}]{pastor2002physics}%
  \BibitemOpen
  \bibfield  {author} {\bibinfo {author} {\bibfnamefont {S.}~\bibnamefont {Pastor}}, \bibinfo {author} {\bibfnamefont {G.}~\bibnamefont {Raffelt}}, \ and\ \bibinfo {author} {\bibfnamefont {D.~V.}\ \bibnamefont {Semikoz}},\ }\href {\doibase 10.1103/PhysRevD.65.053011} {\bibfield  {journal} {\bibinfo  {journal} {Phys. Rev. D}\ }\textbf {\bibinfo {volume} {65}},\ \bibinfo {pages} {053011} (\bibinfo {year} {2002})}\BibitemShut {NoStop}%
\bibitem [{\citenamefont {Duan}\ \emph {et~al.}(2006)\citenamefont {Duan}, \citenamefont {Fuller},\ and\ \citenamefont {Qian}}]{Duan:2005cp}%
  \BibitemOpen
  \bibfield  {author} {\bibinfo {author} {\bibfnamefont {H.}~\bibnamefont {Duan}}, \bibinfo {author} {\bibfnamefont {G.~M.}\ \bibnamefont {Fuller}}, \ and\ \bibinfo {author} {\bibfnamefont {Y.-Z.}\ \bibnamefont {Qian}},\ }\href {\doibase 10.1103/PhysRevD.74.123004} {\bibfield  {journal} {\bibinfo  {journal} {Phys. Rev. D}\ }\textbf {\bibinfo {volume} {74}},\ \bibinfo {pages} {123004} (\bibinfo {year} {2006})},\ \Eprint {http://arxiv.org/abs/astro-ph/0511275} {arXiv:astro-ph/0511275} \BibitemShut {NoStop}%
\bibitem [{\citenamefont {Duan}\ \emph {et~al.}(2010)\citenamefont {Duan}, \citenamefont {Fuller},\ and\ \citenamefont {Qian}}]{annurev:/content/journals/10.1146/annurev.nucl.012809.104524}%
  \BibitemOpen
  \bibfield  {author} {\bibinfo {author} {\bibfnamefont {H.}~\bibnamefont {Duan}}, \bibinfo {author} {\bibfnamefont {G.~M.}\ \bibnamefont {Fuller}}, \ and\ \bibinfo {author} {\bibfnamefont {Y.-Z.}\ \bibnamefont {Qian}},\ }\href {\doibase https://doi.org/10.1146/annurev.nucl.012809.104524} {\bibfield  {journal} {\bibinfo  {journal} {Annual Review of Nuclear and Particle Science}\ }\textbf {\bibinfo {volume} {60}},\ \bibinfo {pages} {569} (\bibinfo {year} {2010})}\BibitemShut {NoStop}%
\bibitem [{\citenamefont {Chakraborty}\ \emph {et~al.}(2016)\citenamefont {Chakraborty}, \citenamefont {Hansen}, \citenamefont {Izaguirre},\ and\ \citenamefont {Raffelt}}]{CHAKRABORTY2016366}%
  \BibitemOpen
  \bibfield  {author} {\bibinfo {author} {\bibfnamefont {S.}~\bibnamefont {Chakraborty}}, \bibinfo {author} {\bibfnamefont {R.}~\bibnamefont {Hansen}}, \bibinfo {author} {\bibfnamefont {I.}~\bibnamefont {Izaguirre}}, \ and\ \bibinfo {author} {\bibfnamefont {G.}~\bibnamefont {Raffelt}},\ }\href {\doibase https://doi.org/10.1016/j.nuclphysb.2016.02.012} {\bibfield  {journal} {\bibinfo  {journal} {Nuclear Physics B}\ }\textbf {\bibinfo {volume} {908}},\ \bibinfo {pages} {366} (\bibinfo {year} {2016})}\BibitemShut {NoStop}%
\bibitem [{\citenamefont {Volpe}(2024)}]{RevModPhys.96.025004}%
  \BibitemOpen
  \bibfield  {author} {\bibinfo {author} {\bibfnamefont {M.~C.}\ \bibnamefont {Volpe}},\ }\href {\doibase 10.1103/RevModPhys.96.025004} {\bibfield  {journal} {\bibinfo  {journal} {Rev. Mod. Phys.}\ }\textbf {\bibinfo {volume} {96}},\ \bibinfo {pages} {025004} (\bibinfo {year} {2024})}\BibitemShut {NoStop}%
\bibitem [{\citenamefont {Bell}\ \emph {et~al.}(2003)\citenamefont {Bell}, \citenamefont {Rawlinson},\ and\ \citenamefont {Sawyer}}]{Bell:2003mg}%
  \BibitemOpen
  \bibfield  {author} {\bibinfo {author} {\bibfnamefont {N.~F.}\ \bibnamefont {Bell}}, \bibinfo {author} {\bibfnamefont {A.~A.}\ \bibnamefont {Rawlinson}}, \ and\ \bibinfo {author} {\bibfnamefont {R.~F.}\ \bibnamefont {Sawyer}},\ }\href {\doibase 10.1016/j.physletb.2003.08.035} {\bibfield  {journal} {\bibinfo  {journal} {Phys. Lett. B}\ }\textbf {\bibinfo {volume} {573}},\ \bibinfo {pages} {86} (\bibinfo {year} {2003})}\BibitemShut {NoStop}%
\bibitem [{\citenamefont {Friedland}\ and\ \citenamefont {Lunardini}(2003{\natexlab{a}})}]{Friedland:2003dv}%
  \BibitemOpen
  \bibfield  {author} {\bibinfo {author} {\bibfnamefont {A.}~\bibnamefont {Friedland}}\ and\ \bibinfo {author} {\bibfnamefont {C.}~\bibnamefont {Lunardini}},\ }\href {\doibase 10.1103/PhysRevD.68.013007} {\bibfield  {journal} {\bibinfo  {journal} {Phys. Rev. D}\ }\textbf {\bibinfo {volume} {68}},\ \bibinfo {pages} {013007} (\bibinfo {year} {2003}{\natexlab{a}})}\BibitemShut {NoStop}%
\bibitem [{\citenamefont {Friedland}\ and\ \citenamefont {Lunardini}(2003{\natexlab{b}})}]{Friedland:2003eh}%
  \BibitemOpen
  \bibfield  {author} {\bibinfo {author} {\bibfnamefont {A.}~\bibnamefont {Friedland}}\ and\ \bibinfo {author} {\bibfnamefont {C.}~\bibnamefont {Lunardini}},\ }\href {\doibase 10.1088/1126-6708/2003/10/043} {\bibfield  {journal} {\bibinfo  {journal} {JHEP}\ }\textbf {\bibinfo {volume} {10}},\ \bibinfo {pages} {043} (\bibinfo {year} {2003}{\natexlab{b}})}\BibitemShut {NoStop}%
\bibitem [{\citenamefont {Friedland}\ \emph {et~al.}(2006)\citenamefont {Friedland}, \citenamefont {McKellar},\ and\ \citenamefont {Okuniewicz}}]{Friedland:2006ke}%
  \BibitemOpen
  \bibfield  {author} {\bibinfo {author} {\bibfnamefont {A.}~\bibnamefont {Friedland}}, \bibinfo {author} {\bibfnamefont {B.~H.~J.}\ \bibnamefont {McKellar}}, \ and\ \bibinfo {author} {\bibfnamefont {I.}~\bibnamefont {Okuniewicz}},\ }\href {\doibase 10.1103/PhysRevD.73.093002} {\bibfield  {journal} {\bibinfo  {journal} {Phys. Rev. D}\ }\textbf {\bibinfo {volume} {73}},\ \bibinfo {pages} {093002} (\bibinfo {year} {2006})}\BibitemShut {NoStop}%
\bibitem [{\citenamefont {McKellar}\ \emph {et~al.}(2009)\citenamefont {McKellar}, \citenamefont {Okuniewicz},\ and\ \citenamefont {Quach}}]{McKellar:2009py}%
  \BibitemOpen
  \bibfield  {author} {\bibinfo {author} {\bibfnamefont {B.~H.~J.}\ \bibnamefont {McKellar}}, \bibinfo {author} {\bibfnamefont {I.}~\bibnamefont {Okuniewicz}}, \ and\ \bibinfo {author} {\bibfnamefont {J.}~\bibnamefont {Quach}},\ }\href {\doibase 10.1103/PhysRevD.80.013011} {\bibfield  {journal} {\bibinfo  {journal} {Phys. Rev. D}\ }\textbf {\bibinfo {volume} {80}},\ \bibinfo {pages} {013011} (\bibinfo {year} {2009})}\BibitemShut {NoStop}%
\bibitem [{\citenamefont {Balantekin}\ and\ \citenamefont {Pehlivan}(2007)}]{Balantekin:2006tg}%
  \BibitemOpen
  \bibfield  {author} {\bibinfo {author} {\bibfnamefont {A.}~\bibnamefont {Balantekin}}\ and\ \bibinfo {author} {\bibfnamefont {Y.}~\bibnamefont {Pehlivan}},\ }\href {\doibase 10.1088/0954-3899/34/1/004} {\bibfield  {journal} {\bibinfo  {journal} {J. Phys. G}\ }\textbf {\bibinfo {volume} {34}},\ \bibinfo {pages} {47} (\bibinfo {year} {2007})},\ \Eprint {http://arxiv.org/abs/astro-ph/0607527} {arXiv:astro-ph/0607527} \BibitemShut {NoStop}%
\bibitem [{\citenamefont {Pehlivan}\ \emph {et~al.}(2011)\citenamefont {Pehlivan}, \citenamefont {Balantekin}, \citenamefont {Kajino},\ and\ \citenamefont {Yoshida}}]{Pehlivan:2011hp}%
  \BibitemOpen
  \bibfield  {author} {\bibinfo {author} {\bibfnamefont {Y.}~\bibnamefont {Pehlivan}}, \bibinfo {author} {\bibfnamefont {A.~B.}\ \bibnamefont {Balantekin}}, \bibinfo {author} {\bibfnamefont {T.}~\bibnamefont {Kajino}}, \ and\ \bibinfo {author} {\bibfnamefont {T.}~\bibnamefont {Yoshida}},\ }\href {\doibase 10.1103/PhysRevD.84.065008} {\bibfield  {journal} {\bibinfo  {journal} {Phys. Rev. D}\ }\textbf {\bibinfo {volume} {84}},\ \bibinfo {pages} {065008} (\bibinfo {year} {2011})}\BibitemShut {NoStop}%
\bibitem [{\citenamefont {Pehlivan}\ \emph {et~al.}(2014)\citenamefont {Pehlivan}, \citenamefont {Balantekin},\ and\ \citenamefont {Kajino}}]{Pehlivan:2014zua}%
  \BibitemOpen
  \bibfield  {author} {\bibinfo {author} {\bibfnamefont {Y.}~\bibnamefont {Pehlivan}}, \bibinfo {author} {\bibfnamefont {A.~B.}\ \bibnamefont {Balantekin}}, \ and\ \bibinfo {author} {\bibfnamefont {T.}~\bibnamefont {Kajino}},\ }\href {\doibase 10.1103/PhysRevD.90.065011} {\bibfield  {journal} {\bibinfo  {journal} {Phys. Rev. D}\ }\textbf {\bibinfo {volume} {90}},\ \bibinfo {pages} {065011} (\bibinfo {year} {2014})}\BibitemShut {NoStop}%
\bibitem [{\citenamefont {Birol}\ \emph {et~al.}(2018)\citenamefont {Birol}, \citenamefont {Pehlivan}, \citenamefont {Balantekin},\ and\ \citenamefont {Kajino}}]{Birol:2018qhx}%
  \BibitemOpen
  \bibfield  {author} {\bibinfo {author} {\bibfnamefont {S.}~\bibnamefont {Birol}}, \bibinfo {author} {\bibfnamefont {Y.}~\bibnamefont {Pehlivan}}, \bibinfo {author} {\bibfnamefont {A.}~\bibnamefont {Balantekin}}, \ and\ \bibinfo {author} {\bibfnamefont {T.}~\bibnamefont {Kajino}},\ }\href {\doibase 10.1103/PhysRevD.98.083002} {\bibfield  {journal} {\bibinfo  {journal} {Phys. Rev. D}\ }\textbf {\bibinfo {volume} {98}},\ \bibinfo {pages} {083002} (\bibinfo {year} {2018})},\ \Eprint {http://arxiv.org/abs/1805.11767} {arXiv:1805.11767 [astro-ph.HE]} \BibitemShut {NoStop}%
\bibitem [{\citenamefont {Patwardhan}\ \emph {et~al.}(2019)\citenamefont {Patwardhan}, \citenamefont {Cervia},\ and\ \citenamefont {Baha~Balantekin}}]{Patwardhan:2019zta}%
  \BibitemOpen
  \bibfield  {author} {\bibinfo {author} {\bibfnamefont {A.~V.}\ \bibnamefont {Patwardhan}}, \bibinfo {author} {\bibfnamefont {M.~J.}\ \bibnamefont {Cervia}}, \ and\ \bibinfo {author} {\bibfnamefont {A.}~\bibnamefont {Baha~Balantekin}},\ }\href {\doibase 10.1103/PhysRevD.99.123013} {\bibfield  {journal} {\bibinfo  {journal} {Phys. Rev. D}\ }\textbf {\bibinfo {volume} {99}},\ \bibinfo {pages} {123013} (\bibinfo {year} {2019})}\BibitemShut {NoStop}%
\bibitem [{\citenamefont {Cervia}\ \emph {et~al.}(2019)\citenamefont {Cervia}, \citenamefont {Patwardhan}, \citenamefont {Balantekin}, \citenamefont {Coppersmith},\ and\ \citenamefont {Johnson}}]{Cervia:2019res}%
  \BibitemOpen
  \bibfield  {author} {\bibinfo {author} {\bibfnamefont {M.~J.}\ \bibnamefont {Cervia}}, \bibinfo {author} {\bibfnamefont {A.~V.}\ \bibnamefont {Patwardhan}}, \bibinfo {author} {\bibfnamefont {A.~B.}\ \bibnamefont {Balantekin}}, \bibinfo {author} {\bibfnamefont {t.~S.~N.}\ \bibnamefont {Coppersmith}}, \ and\ \bibinfo {author} {\bibfnamefont {C.~W.}\ \bibnamefont {Johnson}},\ }\href {\doibase 10.1103/PhysRevD.100.083001} {\bibfield  {journal} {\bibinfo  {journal} {Phys. Rev. D}\ }\textbf {\bibinfo {volume} {100}},\ \bibinfo {pages} {083001} (\bibinfo {year} {2019})}\BibitemShut {NoStop}%
\bibitem [{\citenamefont {Rrapaj}(2020)}]{Rrapaj:2019pxz}%
  \BibitemOpen
  \bibfield  {author} {\bibinfo {author} {\bibfnamefont {E.}~\bibnamefont {Rrapaj}},\ }\href {\doibase 10.1103/PhysRevC.101.065805} {\bibfield  {journal} {\bibinfo  {journal} {Phys. Rev. C}\ }\textbf {\bibinfo {volume} {101}},\ \bibinfo {pages} {065805} (\bibinfo {year} {2020})}\BibitemShut {NoStop}%
\bibitem [{\citenamefont {Roggero}(2021{\natexlab{a}})}]{Roggero2021a}%
  \BibitemOpen
  \bibfield  {author} {\bibinfo {author} {\bibfnamefont {A.}~\bibnamefont {Roggero}},\ }\href {\doibase 10.1103/PhysRevD.104.103016} {\bibfield  {journal} {\bibinfo  {journal} {Phys. Rev. D}\ }\textbf {\bibinfo {volume} {104}},\ \bibinfo {pages} {103016} (\bibinfo {year} {2021}{\natexlab{a}})}\BibitemShut {NoStop}%
\bibitem [{\citenamefont {Roggero}(2021{\natexlab{b}})}]{Roggero2021b}%
  \BibitemOpen
  \bibfield  {author} {\bibinfo {author} {\bibfnamefont {A.}~\bibnamefont {Roggero}},\ }\href {\doibase 10.1103/PhysRevD.104.123023} {\bibfield  {journal} {\bibinfo  {journal} {Phys. Rev. D}\ }\textbf {\bibinfo {volume} {104}},\ \bibinfo {pages} {123023} (\bibinfo {year} {2021}{\natexlab{b}})}\BibitemShut {NoStop}%
\bibitem [{\citenamefont {Xiong}(2022)}]{Xiong:2021evk}%
  \BibitemOpen
  \bibfield  {author} {\bibinfo {author} {\bibfnamefont {Z.}~\bibnamefont {Xiong}},\ }\href {\doibase 10.1103/PhysRevD.105.103002} {\bibfield  {journal} {\bibinfo  {journal} {Phys. Rev. D}\ }\textbf {\bibinfo {volume} {105}},\ \bibinfo {pages} {103002} (\bibinfo {year} {2022})},\ \Eprint {http://arxiv.org/abs/2111.00437} {arXiv:2111.00437 [astro-ph.HE]} \BibitemShut {NoStop}%
\bibitem [{\citenamefont {Martin}\ \emph {et~al.}(2022)\citenamefont {Martin}, \citenamefont {Roggero}, \citenamefont {Duan}, \citenamefont {Carlson},\ and\ \citenamefont {Cirigliano}}]{Martin:2021bri}%
  \BibitemOpen
  \bibfield  {author} {\bibinfo {author} {\bibfnamefont {J.~D.}\ \bibnamefont {Martin}}, \bibinfo {author} {\bibfnamefont {A.}~\bibnamefont {Roggero}}, \bibinfo {author} {\bibfnamefont {H.}~\bibnamefont {Duan}}, \bibinfo {author} {\bibfnamefont {J.}~\bibnamefont {Carlson}}, \ and\ \bibinfo {author} {\bibfnamefont {V.}~\bibnamefont {Cirigliano}},\ }\href {\doibase 10.1103/PhysRevD.105.083020} {\bibfield  {journal} {\bibinfo  {journal} {Phys. Rev. D}\ }\textbf {\bibinfo {volume} {105}},\ \bibinfo {pages} {083020} (\bibinfo {year} {2022})},\ \Eprint {http://arxiv.org/abs/2112.12686} {arXiv:2112.12686 [hep-ph]} \BibitemShut {NoStop}%
\bibitem [{\citenamefont {Patwardhan}\ \emph {et~al.}(2021)\citenamefont {Patwardhan}, \citenamefont {Cervia},\ and\ \citenamefont {Balantekin}}]{Patwardhan:2021rej}%
  \BibitemOpen
  \bibfield  {author} {\bibinfo {author} {\bibfnamefont {A.~V.}\ \bibnamefont {Patwardhan}}, \bibinfo {author} {\bibfnamefont {M.~J.}\ \bibnamefont {Cervia}}, \ and\ \bibinfo {author} {\bibfnamefont {A.~B.}\ \bibnamefont {Balantekin}},\ }\href {\doibase 10.1103/PhysRevD.104.123035} {\bibfield  {journal} {\bibinfo  {journal} {Phys. Rev. D}\ }\textbf {\bibinfo {volume} {104}},\ \bibinfo {pages} {123035} (\bibinfo {year} {2021})}\BibitemShut {NoStop}%
\bibitem [{\citenamefont {Roggero}\ \emph {et~al.}(2022)\citenamefont {Roggero}, \citenamefont {Rrapaj},\ and\ \citenamefont {Xiong}}]{Roggero2022}%
  \BibitemOpen
  \bibfield  {author} {\bibinfo {author} {\bibfnamefont {A.}~\bibnamefont {Roggero}}, \bibinfo {author} {\bibfnamefont {E.}~\bibnamefont {Rrapaj}}, \ and\ \bibinfo {author} {\bibfnamefont {Z.}~\bibnamefont {Xiong}},\ }\href {\doibase 10.1103/PhysRevD.106.043022} {\bibfield  {journal} {\bibinfo  {journal} {Phys. Rev. D}\ }\textbf {\bibinfo {volume} {106}},\ \bibinfo {pages} {043022} (\bibinfo {year} {2022})}\BibitemShut {NoStop}%
\bibitem [{\citenamefont {Cervia}\ \emph {et~al.}(2022)\citenamefont {Cervia}, \citenamefont {Siwach}, \citenamefont {Patwardhan}, \citenamefont {Balantekin}, \citenamefont {Coppersmith},\ and\ \citenamefont {Johnson}}]{Cervia:2022pro}%
  \BibitemOpen
  \bibfield  {author} {\bibinfo {author} {\bibfnamefont {M.~J.}\ \bibnamefont {Cervia}}, \bibinfo {author} {\bibfnamefont {P.}~\bibnamefont {Siwach}}, \bibinfo {author} {\bibfnamefont {A.~V.}\ \bibnamefont {Patwardhan}}, \bibinfo {author} {\bibfnamefont {A.~B.}\ \bibnamefont {Balantekin}}, \bibinfo {author} {\bibfnamefont {S.~N.}\ \bibnamefont {Coppersmith}}, \ and\ \bibinfo {author} {\bibfnamefont {C.~W.}\ \bibnamefont {Johnson}},\ }\href {\doibase 10.1103/PhysRevD.105.123025} {\bibfield  {journal} {\bibinfo  {journal} {Phys. Rev. D}\ }\textbf {\bibinfo {volume} {105}},\ \bibinfo {pages} {123025} (\bibinfo {year} {2022})}\BibitemShut {NoStop}%
\bibitem [{\citenamefont {Illa}\ and\ \citenamefont {Savage}(2023)}]{Illa:2022zgu}%
  \BibitemOpen
  \bibfield  {author} {\bibinfo {author} {\bibfnamefont {M.}~\bibnamefont {Illa}}\ and\ \bibinfo {author} {\bibfnamefont {M.~J.}\ \bibnamefont {Savage}},\ }\href {\doibase 10.1103/PhysRevLett.130.221003} {\bibfield  {journal} {\bibinfo  {journal} {Phys. Rev. Lett.}\ }\textbf {\bibinfo {volume} {130}},\ \bibinfo {pages} {221003} (\bibinfo {year} {2023})}\BibitemShut {NoStop}%
\bibitem [{\citenamefont {Lacroix}\ \emph {et~al.}(2022)\citenamefont {Lacroix}, \citenamefont {Balantekin}, \citenamefont {Cervia}, \citenamefont {Patwardhan},\ and\ \citenamefont {Siwach}}]{Lacroix:2022krq}%
  \BibitemOpen
  \bibfield  {author} {\bibinfo {author} {\bibfnamefont {D.}~\bibnamefont {Lacroix}}, \bibinfo {author} {\bibfnamefont {A.~B.}\ \bibnamefont {Balantekin}}, \bibinfo {author} {\bibfnamefont {M.~J.}\ \bibnamefont {Cervia}}, \bibinfo {author} {\bibfnamefont {A.~V.}\ \bibnamefont {Patwardhan}}, \ and\ \bibinfo {author} {\bibfnamefont {P.}~\bibnamefont {Siwach}},\ }\href {\doibase 10.1103/PhysRevD.106.123006} {\bibfield  {journal} {\bibinfo  {journal} {Phys. Rev. D}\ }\textbf {\bibinfo {volume} {106}},\ \bibinfo {pages} {123006} (\bibinfo {year} {2022})}\BibitemShut {NoStop}%
\bibitem [{\citenamefont {Siwach}\ \emph {et~al.}(2022)\citenamefont {Siwach}, \citenamefont {Suliga},\ and\ \citenamefont {Balantekin}}]{Siwach:2022xhx}%
  \BibitemOpen
  \bibfield  {author} {\bibinfo {author} {\bibfnamefont {P.}~\bibnamefont {Siwach}}, \bibinfo {author} {\bibfnamefont {A.~M.}\ \bibnamefont {Suliga}}, \ and\ \bibinfo {author} {\bibfnamefont {A.~B.}\ \bibnamefont {Balantekin}},\ }\href@noop {} {\  (\bibinfo {year} {2022})},\ \Eprint {http://arxiv.org/abs/2211.07678} {arXiv:2211.07678 [hep-ph]} \BibitemShut {NoStop}%
\bibitem [{\citenamefont {Bhaskar}\ \emph {et~al.}(2024)\citenamefont {Bhaskar}, \citenamefont {Roggero},\ and\ \citenamefont {Savage}}]{PhysRevC.110.045801}%
  \BibitemOpen
  \bibfield  {author} {\bibinfo {author} {\bibfnamefont {R.}~\bibnamefont {Bhaskar}}, \bibinfo {author} {\bibfnamefont {A.}~\bibnamefont {Roggero}}, \ and\ \bibinfo {author} {\bibfnamefont {M.~J.}\ \bibnamefont {Savage}},\ }\href {\doibase 10.1103/PhysRevC.110.045801} {\bibfield  {journal} {\bibinfo  {journal} {Phys. Rev. C}\ }\textbf {\bibinfo {volume} {110}},\ \bibinfo {pages} {045801} (\bibinfo {year} {2024})}\BibitemShut {NoStop}%
\bibitem [{\citenamefont {Martin}\ \emph {et~al.}(2023{\natexlab{a}})\citenamefont {Martin}, \citenamefont {Neill}, \citenamefont {Roggero}, \citenamefont {Duan},\ and\ \citenamefont {Carlson}}]{Martin:2023gbo}%
  \BibitemOpen
  \bibfield  {author} {\bibinfo {author} {\bibfnamefont {J.~D.}\ \bibnamefont {Martin}}, \bibinfo {author} {\bibfnamefont {D.}~\bibnamefont {Neill}}, \bibinfo {author} {\bibfnamefont {A.}~\bibnamefont {Roggero}}, \bibinfo {author} {\bibfnamefont {H.}~\bibnamefont {Duan}}, \ and\ \bibinfo {author} {\bibfnamefont {J.}~\bibnamefont {Carlson}},\ }\href {\doibase 10.1103/PhysRevD.108.123010} {\bibfield  {journal} {\bibinfo  {journal} {Phys. Rev. D}\ }\textbf {\bibinfo {volume} {108}},\ \bibinfo {pages} {123010} (\bibinfo {year} {2023}{\natexlab{a}})}\BibitemShut {NoStop}%
\bibitem [{\citenamefont {Martin}\ \emph {et~al.}(2023{\natexlab{b}})\citenamefont {Martin}, \citenamefont {Roggero}, \citenamefont {Duan},\ and\ \citenamefont {Carlson}}]{Martin:2023ljq}%
  \BibitemOpen
  \bibfield  {author} {\bibinfo {author} {\bibfnamefont {J.~D.}\ \bibnamefont {Martin}}, \bibinfo {author} {\bibfnamefont {A.}~\bibnamefont {Roggero}}, \bibinfo {author} {\bibfnamefont {H.}~\bibnamefont {Duan}}, \ and\ \bibinfo {author} {\bibfnamefont {J.}~\bibnamefont {Carlson}},\ }\href@noop {} {\  (\bibinfo {year} {2023}{\natexlab{b}})},\ \Eprint {http://arxiv.org/abs/2301.07049} {arXiv:2301.07049 [hep-ph]} \BibitemShut {NoStop}%
\bibitem [{\citenamefont {Kost}\ \emph {et~al.}(2024)\citenamefont {Kost}, \citenamefont {Johns},\ and\ \citenamefont {Duan}}]{PhysRevD.109.103037}%
  \BibitemOpen
  \bibfield  {author} {\bibinfo {author} {\bibfnamefont {A.}~\bibnamefont {Kost}}, \bibinfo {author} {\bibfnamefont {L.}~\bibnamefont {Johns}}, \ and\ \bibinfo {author} {\bibfnamefont {H.}~\bibnamefont {Duan}},\ }\href {\doibase 10.1103/PhysRevD.109.103037} {\bibfield  {journal} {\bibinfo  {journal} {Phys. Rev. D}\ }\textbf {\bibinfo {volume} {109}},\ \bibinfo {pages} {103037} (\bibinfo {year} {2024})}\BibitemShut {NoStop}%
\bibitem [{\citenamefont {Lacroix}\ \emph {et~al.}(2024)\citenamefont {Lacroix}, \citenamefont {Bauge}, \citenamefont {Yilmaz}, \citenamefont {Mangin-Brinet}, \citenamefont {Roggero},\ and\ \citenamefont {Balantekin}}]{PhysRevD.110.103027}%
  \BibitemOpen
  \bibfield  {author} {\bibinfo {author} {\bibfnamefont {D.}~\bibnamefont {Lacroix}}, \bibinfo {author} {\bibfnamefont {A.}~\bibnamefont {Bauge}}, \bibinfo {author} {\bibfnamefont {B.}~\bibnamefont {Yilmaz}}, \bibinfo {author} {\bibfnamefont {M.}~\bibnamefont {Mangin-Brinet}}, \bibinfo {author} {\bibfnamefont {A.}~\bibnamefont {Roggero}}, \ and\ \bibinfo {author} {\bibfnamefont {A.~B.}\ \bibnamefont {Balantekin}},\ }\href {\doibase 10.1103/PhysRevD.110.103027} {\bibfield  {journal} {\bibinfo  {journal} {Phys. Rev. D}\ }\textbf {\bibinfo {volume} {110}},\ \bibinfo {pages} {103027} (\bibinfo {year} {2024})}\BibitemShut {NoStop}%
\bibitem [{\citenamefont {Cirigliano}\ \emph {et~al.}(2024)\citenamefont {Cirigliano}, \citenamefont {Sen},\ and\ \citenamefont {Yamauchi}}]{Cirigliano:2024}%
  \BibitemOpen
  \bibfield  {author} {\bibinfo {author} {\bibfnamefont {V.}~\bibnamefont {Cirigliano}}, \bibinfo {author} {\bibfnamefont {S.}~\bibnamefont {Sen}}, \ and\ \bibinfo {author} {\bibfnamefont {Y.}~\bibnamefont {Yamauchi}},\ }\href {\doibase 10.1103/PhysRevD.110.123028} {\bibfield  {journal} {\bibinfo  {journal} {Phys. Rev. D}\ }\textbf {\bibinfo {volume} {110}},\ \bibinfo {pages} {123028} (\bibinfo {year} {2024})}\BibitemShut {NoStop}%
\bibitem [{\citenamefont {Laraib}\ and\ \citenamefont {Richers}(2025)}]{d6w8-7j9s}%
  \BibitemOpen
  \bibfield  {author} {\bibinfo {author} {\bibfnamefont {Z.}~\bibnamefont {Laraib}}\ and\ \bibinfo {author} {\bibfnamefont {S.}~\bibnamefont {Richers}},\ }\href {\doibase 10.1103/d6w8-7j9s} {\bibfield  {journal} {\bibinfo  {journal} {Phys. Rev. D}\ }\textbf {\bibinfo {volume} {112}},\ \bibinfo {pages} {L101304} (\bibinfo {year} {2025})}\BibitemShut {NoStop}%
\bibitem [{\citenamefont {Chernyshev}\ \emph {et~al.}(2025)\citenamefont {Chernyshev}, \citenamefont {Robin},\ and\ \citenamefont {Savage}}]{PhysRevResearch.7.023228}%
  \BibitemOpen
  \bibfield  {author} {\bibinfo {author} {\bibfnamefont {I.}~\bibnamefont {Chernyshev}}, \bibinfo {author} {\bibfnamefont {C.~E.~P.}\ \bibnamefont {Robin}}, \ and\ \bibinfo {author} {\bibfnamefont {M.~J.}\ \bibnamefont {Savage}},\ }\href {\doibase 10.1103/PhysRevResearch.7.023228} {\bibfield  {journal} {\bibinfo  {journal} {Phys. Rev. Res.}\ }\textbf {\bibinfo {volume} {7}},\ \bibinfo {pages} {023228} (\bibinfo {year} {2025})}\BibitemShut {NoStop}%
\bibitem [{\citenamefont {Mangin-Brinet}\ \emph {et~al.}(2025)\citenamefont {Mangin-Brinet}, \citenamefont {Bauge},\ and\ \citenamefont {Lacroix}}]{manginbrinet2025threeflavorneutrinooscillationsusing}%
  \BibitemOpen
  \bibfield  {author} {\bibinfo {author} {\bibfnamefont {M.}~\bibnamefont {Mangin-Brinet}}, \bibinfo {author} {\bibfnamefont {A.}~\bibnamefont {Bauge}}, \ and\ \bibinfo {author} {\bibfnamefont {D.}~\bibnamefont {Lacroix}},\ }\href {https://arxiv.org/abs/2507.18482} {\enquote {\bibinfo {title} {Three-flavor neutrino oscillations using the phase space approach},}\ } (\bibinfo {year} {2025}),\ \Eprint {http://arxiv.org/abs/2507.18482} {arXiv:2507.18482 [hep-ph]} \BibitemShut {NoStop}%
\bibitem [{\citenamefont {Hall}\ \emph {et~al.}(2021)\citenamefont {Hall}, \citenamefont {Roggero}, \citenamefont {Baroni},\ and\ \citenamefont {Carlson}}]{Hall:2021rbv}%
  \BibitemOpen
  \bibfield  {author} {\bibinfo {author} {\bibfnamefont {B.}~\bibnamefont {Hall}}, \bibinfo {author} {\bibfnamefont {A.}~\bibnamefont {Roggero}}, \bibinfo {author} {\bibfnamefont {A.}~\bibnamefont {Baroni}}, \ and\ \bibinfo {author} {\bibfnamefont {J.}~\bibnamefont {Carlson}},\ }\href {\doibase 10.1103/PhysRevD.104.063009} {\bibfield  {journal} {\bibinfo  {journal} {Phys. Rev. D}\ }\textbf {\bibinfo {volume} {104}},\ \bibinfo {pages} {063009} (\bibinfo {year} {2021})}\BibitemShut {NoStop}%
\bibitem [{\citenamefont {Yeter-Aydeniz}\ \emph {et~al.}(2022)\citenamefont {Yeter-Aydeniz}, \citenamefont {Bangar}, \citenamefont {Siopsis},\ and\ \citenamefont {Pooser}}]{Yeter-Aydeniz:2021olz}%
  \BibitemOpen
  \bibfield  {author} {\bibinfo {author} {\bibfnamefont {K.}~\bibnamefont {Yeter-Aydeniz}}, \bibinfo {author} {\bibfnamefont {S.}~\bibnamefont {Bangar}}, \bibinfo {author} {\bibfnamefont {G.}~\bibnamefont {Siopsis}}, \ and\ \bibinfo {author} {\bibfnamefont {R.~C.}\ \bibnamefont {Pooser}},\ }\href {\doibase 10.1007/s11128-021-03348-x} {\bibfield  {journal} {\bibinfo  {journal} {Quant. Inf. Proc.}\ }\textbf {\bibinfo {volume} {21}},\ \bibinfo {pages} {84} (\bibinfo {year} {2022})}\BibitemShut {NoStop}%
\bibitem [{\citenamefont {Illa}\ and\ \citenamefont {Savage}(2022)}]{PhysRevA.106.052605}%
  \BibitemOpen
  \bibfield  {author} {\bibinfo {author} {\bibfnamefont {M.}~\bibnamefont {Illa}}\ and\ \bibinfo {author} {\bibfnamefont {M.~J.}\ \bibnamefont {Savage}},\ }\href {\doibase 10.1103/PhysRevA.106.052605} {\bibfield  {journal} {\bibinfo  {journal} {Phys. Rev. A}\ }\textbf {\bibinfo {volume} {106}},\ \bibinfo {pages} {052605} (\bibinfo {year} {2022})}\BibitemShut {NoStop}%
\bibitem [{\citenamefont {Amitrano}\ \emph {et~al.}(2023)\citenamefont {Amitrano}, \citenamefont {Roggero}, \citenamefont {Luchi}, \citenamefont {Turro}, \citenamefont {Vespucci},\ and\ \citenamefont {Pederiva}}]{Amitrano2023}%
  \BibitemOpen
  \bibfield  {author} {\bibinfo {author} {\bibfnamefont {V.}~\bibnamefont {Amitrano}}, \bibinfo {author} {\bibfnamefont {A.}~\bibnamefont {Roggero}}, \bibinfo {author} {\bibfnamefont {P.}~\bibnamefont {Luchi}}, \bibinfo {author} {\bibfnamefont {F.}~\bibnamefont {Turro}}, \bibinfo {author} {\bibfnamefont {L.}~\bibnamefont {Vespucci}}, \ and\ \bibinfo {author} {\bibfnamefont {F.}~\bibnamefont {Pederiva}},\ }\href {\doibase 10.1103/PhysRevD.107.023007} {\bibfield  {journal} {\bibinfo  {journal} {Phys. Rev. D}\ }\textbf {\bibinfo {volume} {107}},\ \bibinfo {pages} {023007} (\bibinfo {year} {2023})}\BibitemShut {NoStop}%
\bibitem [{\citenamefont {Siwach}\ \emph {et~al.}(2023)\citenamefont {Siwach}, \citenamefont {Harrison},\ and\ \citenamefont {Balantekin}}]{Siwach:2023wzy}%
  \BibitemOpen
  \bibfield  {author} {\bibinfo {author} {\bibfnamefont {P.}~\bibnamefont {Siwach}}, \bibinfo {author} {\bibfnamefont {K.}~\bibnamefont {Harrison}}, \ and\ \bibinfo {author} {\bibfnamefont {A.~B.}\ \bibnamefont {Balantekin}},\ }\href {\doibase 10.1103/PhysRevD.108.083039} {\bibfield  {journal} {\bibinfo  {journal} {Phys. Rev. D}\ }\textbf {\bibinfo {volume} {108}},\ \bibinfo {pages} {083039} (\bibinfo {year} {2023})}\BibitemShut {NoStop}%
\bibitem [{\citenamefont {Turro}\ \emph {et~al.}(2025)\citenamefont {Turro}, \citenamefont {Chernyshev}, \citenamefont {Bhaskar},\ and\ \citenamefont {Illa}}]{Turro:2025}%
  \BibitemOpen
  \bibfield  {author} {\bibinfo {author} {\bibfnamefont {F.}~\bibnamefont {Turro}}, \bibinfo {author} {\bibfnamefont {I.~A.}\ \bibnamefont {Chernyshev}}, \bibinfo {author} {\bibfnamefont {R.}~\bibnamefont {Bhaskar}}, \ and\ \bibinfo {author} {\bibfnamefont {M.}~\bibnamefont {Illa}},\ }\href {\doibase 10.1103/PhysRevD.111.043038} {\bibfield  {journal} {\bibinfo  {journal} {Phys. Rev. D}\ }\textbf {\bibinfo {volume} {111}},\ \bibinfo {pages} {043038} (\bibinfo {year} {2025})}\BibitemShut {NoStop}%
\bibitem [{\citenamefont {Chernyshev}(2025)}]{PhysRevD.111.043017}%
  \BibitemOpen
  \bibfield  {author} {\bibinfo {author} {\bibfnamefont {I.~A.}\ \bibnamefont {Chernyshev}},\ }\href {\doibase 10.1103/PhysRevD.111.043017} {\bibfield  {journal} {\bibinfo  {journal} {Phys. Rev. D}\ }\textbf {\bibinfo {volume} {111}},\ \bibinfo {pages} {043017} (\bibinfo {year} {2025})}\BibitemShut {NoStop}%
\bibitem [{\citenamefont {Spagnoli}\ \emph {et~al.}(2025)\citenamefont {Spagnoli}, \citenamefont {Goss}, \citenamefont {Roggero}, \citenamefont {Rrapaj}, \citenamefont {Cervia}, \citenamefont {Patwardhan}, \citenamefont {Naik}, \citenamefont {Balantekin}, \citenamefont {Younis}, \citenamefont {Santiago}, \citenamefont {Siddiqi},\ and\ \citenamefont {Aldaihan}}]{gjr1-lf8s}%
  \BibitemOpen
  \bibfield  {author} {\bibinfo {author} {\bibfnamefont {L.}~\bibnamefont {Spagnoli}}, \bibinfo {author} {\bibfnamefont {N.}~\bibnamefont {Goss}}, \bibinfo {author} {\bibfnamefont {A.}~\bibnamefont {Roggero}}, \bibinfo {author} {\bibfnamefont {E.}~\bibnamefont {Rrapaj}}, \bibinfo {author} {\bibfnamefont {M.~J.}\ \bibnamefont {Cervia}}, \bibinfo {author} {\bibfnamefont {A.~V.}\ \bibnamefont {Patwardhan}}, \bibinfo {author} {\bibfnamefont {R.~K.}\ \bibnamefont {Naik}}, \bibinfo {author} {\bibfnamefont {A.~B.}\ \bibnamefont {Balantekin}}, \bibinfo {author} {\bibfnamefont {E.}~\bibnamefont {Younis}}, \bibinfo {author} {\bibfnamefont {D.~I.}\ \bibnamefont {Santiago}}, \bibinfo {author} {\bibfnamefont {I.}~\bibnamefont {Siddiqi}}, \ and\ \bibinfo {author} {\bibfnamefont {S.}~\bibnamefont {Aldaihan}},\ }\href {\doibase 10.1103/gjr1-lf8s} {\bibfield  {journal} {\bibinfo  {journal} {Phys. Rev. D}\ }\textbf {\bibinfo {volume} {111}},\ \bibinfo {pages} {103054} (\bibinfo {year} {2025})}\BibitemShut {NoStop}%
\bibitem [{\citenamefont {Kiss}\ \emph {et~al.}(2025)\citenamefont {Kiss}, \citenamefont {Tavernelli}, \citenamefont {Tacchino}, \citenamefont {Lacroix},\ and\ \citenamefont {Roggero}}]{kiss2025neutrinothermalizationrandomizationquantum}%
  \BibitemOpen
  \bibfield  {author} {\bibinfo {author} {\bibfnamefont {O.}~\bibnamefont {Kiss}}, \bibinfo {author} {\bibfnamefont {I.}~\bibnamefont {Tavernelli}}, \bibinfo {author} {\bibfnamefont {F.}~\bibnamefont {Tacchino}}, \bibinfo {author} {\bibfnamefont {D.}~\bibnamefont {Lacroix}}, \ and\ \bibinfo {author} {\bibfnamefont {A.}~\bibnamefont {Roggero}},\ }\href {https://arxiv.org/abs/2510.24841} {\enquote {\bibinfo {title} {Neutrino thermalization via randomization on a quantum processor},}\ } (\bibinfo {year} {2025}),\ \Eprint {http://arxiv.org/abs/2510.24841} {arXiv:2510.24841 [quant-ph]} \BibitemShut {NoStop}%
\bibitem [{\citenamefont {Martin}\ \emph {et~al.}(2023{\natexlab{c}})\citenamefont {Martin}, \citenamefont {Neill}, \citenamefont {Roggero}, \citenamefont {Duan},\ and\ \citenamefont {Carlson}}]{Martin:2023}%
  \BibitemOpen
  \bibfield  {author} {\bibinfo {author} {\bibfnamefont {J.~D.}\ \bibnamefont {Martin}}, \bibinfo {author} {\bibfnamefont {D.}~\bibnamefont {Neill}}, \bibinfo {author} {\bibfnamefont {A.}~\bibnamefont {Roggero}}, \bibinfo {author} {\bibfnamefont {H.}~\bibnamefont {Duan}}, \ and\ \bibinfo {author} {\bibfnamefont {J.}~\bibnamefont {Carlson}},\ }\href {\doibase 10.1103/PhysRevD.108.123010} {\bibfield  {journal} {\bibinfo  {journal} {Phys. Rev. D}\ }\textbf {\bibinfo {volume} {108}},\ \bibinfo {pages} {123010} (\bibinfo {year} {2023}{\natexlab{c}})}\BibitemShut {NoStop}%
\bibitem [{\citenamefont {Pontecorvo}(1958)}]{Pontecorvo:1957cp}%
  \BibitemOpen
  \bibfield  {author} {\bibinfo {author} {\bibfnamefont {B.}~\bibnamefont {Pontecorvo}},\ }\href@noop {} {\bibfield  {journal} {\bibinfo  {journal} {Sov. Phys. JETP}\ }\textbf {\bibinfo {volume} {6}},\ \bibinfo {pages} {429} (\bibinfo {year} {1958})}\BibitemShut {NoStop}%
\bibitem [{\citenamefont {Maki}\ \emph {et~al.}(1962)\citenamefont {Maki}, \citenamefont {Nakagawa},\ and\ \citenamefont {Sakata}}]{10.1143/PTP.28.870}%
  \BibitemOpen
  \bibfield  {author} {\bibinfo {author} {\bibfnamefont {Z.}~\bibnamefont {Maki}}, \bibinfo {author} {\bibfnamefont {M.}~\bibnamefont {Nakagawa}}, \ and\ \bibinfo {author} {\bibfnamefont {S.}~\bibnamefont {Sakata}},\ }\href {\doibase 10.1143/PTP.28.870} {\bibfield  {journal} {\bibinfo  {journal} {Progress of Theoretical Physics}\ }\textbf {\bibinfo {volume} {28}},\ \bibinfo {pages} {870} (\bibinfo {year} {1962})}\BibitemShut {NoStop}%
\bibitem [{\citenamefont {Hirsch}\ \emph {et~al.}(1982)\citenamefont {Hirsch}, \citenamefont {Sugar}, \citenamefont {Scalapino},\ and\ \citenamefont {Blankenbecler}}]{PhysRevB.26.5033}%
  \BibitemOpen
  \bibfield  {author} {\bibinfo {author} {\bibfnamefont {J.~E.}\ \bibnamefont {Hirsch}}, \bibinfo {author} {\bibfnamefont {R.~L.}\ \bibnamefont {Sugar}}, \bibinfo {author} {\bibfnamefont {D.~J.}\ \bibnamefont {Scalapino}}, \ and\ \bibinfo {author} {\bibfnamefont {R.}~\bibnamefont {Blankenbecler}},\ }\href {\doibase 10.1103/PhysRevB.26.5033} {\bibfield  {journal} {\bibinfo  {journal} {Phys. Rev. B}\ }\textbf {\bibinfo {volume} {26}},\ \bibinfo {pages} {5033} (\bibinfo {year} {1982})}\BibitemShut {NoStop}%
\bibitem [{\citenamefont {Sandvik}\ and\ \citenamefont {Kurkij\"arvi}(1991)}]{PhysRevB.43.5950}%
  \BibitemOpen
  \bibfield  {author} {\bibinfo {author} {\bibfnamefont {A.~W.}\ \bibnamefont {Sandvik}}\ and\ \bibinfo {author} {\bibfnamefont {J.}~\bibnamefont {Kurkij\"arvi}},\ }\href {\doibase 10.1103/PhysRevB.43.5950} {\bibfield  {journal} {\bibinfo  {journal} {Phys. Rev. B}\ }\textbf {\bibinfo {volume} {43}},\ \bibinfo {pages} {5950} (\bibinfo {year} {1991})}\BibitemShut {NoStop}%
\bibitem [{\citenamefont {Fetter}\ and\ \citenamefont {Walecka}(1971)}]{Fetter-Walecka}%
  \BibitemOpen
  \bibfield  {author} {\bibinfo {author} {\bibfnamefont {A.~L.}\ \bibnamefont {Fetter}}\ and\ \bibinfo {author} {\bibfnamefont {J.~D.}\ \bibnamefont {Walecka}},\ }\href@noop {} {\emph {\bibinfo {title} {Quantum Theory of Many-Particle Systems}}}\ (\bibinfo  {publisher} {McGraw-Hill},\ \bibinfo {address} {Boston},\ \bibinfo {year} {1971})\BibitemShut {NoStop}%
\bibitem [{\citenamefont {Margolus}\ and\ \citenamefont {Levitin}(1998)}]{MARGOLUS1998188}%
  \BibitemOpen
  \bibfield  {author} {\bibinfo {author} {\bibfnamefont {N.}~\bibnamefont {Margolus}}\ and\ \bibinfo {author} {\bibfnamefont {L.~B.}\ \bibnamefont {Levitin}},\ }\href {\doibase https://doi.org/10.1016/S0167-2789(98)00054-2} {\bibfield  {journal} {\bibinfo  {journal} {Physica D: Nonlinear Phenomena}\ }\textbf {\bibinfo {volume} {120}},\ \bibinfo {pages} {188} (\bibinfo {year} {1998})},\ \bibinfo {note} {proceedings of the Fourth Workshop on Physics and Consumption}\BibitemShut {NoStop}%
\bibitem [{\citenamefont {Deffner}\ and\ \citenamefont {Campbell}(2017)}]{Deffner_2017}%
  \BibitemOpen
  \bibfield  {author} {\bibinfo {author} {\bibfnamefont {S.}~\bibnamefont {Deffner}}\ and\ \bibinfo {author} {\bibfnamefont {S.}~\bibnamefont {Campbell}},\ }\href {\doibase 10.1088/1751-8121/aa86c6} {\bibfield  {journal} {\bibinfo  {journal} {Journal of Physics A: Mathematical and Theoretical}\ }\textbf {\bibinfo {volume} {50}},\ \bibinfo {pages} {453001} (\bibinfo {year} {2017})}\BibitemShut {NoStop}%
\bibitem [{\citenamefont {Campaioli}\ \emph {et~al.}(2018)\citenamefont {Campaioli}, \citenamefont {Pollock}, \citenamefont {Binder},\ and\ \citenamefont {Modi}}]{PhysRevLett.120.060409}%
  \BibitemOpen
  \bibfield  {author} {\bibinfo {author} {\bibfnamefont {F.}~\bibnamefont {Campaioli}}, \bibinfo {author} {\bibfnamefont {F.~A.}\ \bibnamefont {Pollock}}, \bibinfo {author} {\bibfnamefont {F.~C.}\ \bibnamefont {Binder}}, \ and\ \bibinfo {author} {\bibfnamefont {K.}~\bibnamefont {Modi}},\ }\href {\doibase 10.1103/PhysRevLett.120.060409} {\bibfield  {journal} {\bibinfo  {journal} {Phys. Rev. Lett.}\ }\textbf {\bibinfo {volume} {120}},\ \bibinfo {pages} {060409} (\bibinfo {year} {2018})}\BibitemShut {NoStop}%
\bibitem [{\citenamefont {Neill}\ \emph {et~al.}(2025)\citenamefont {Neill}, \citenamefont {Liu}, \citenamefont {Martin},\ and\ \citenamefont {Roggero}}]{PhysRevResearch.7.023157}%
  \BibitemOpen
  \bibfield  {author} {\bibinfo {author} {\bibfnamefont {D.}~\bibnamefont {Neill}}, \bibinfo {author} {\bibfnamefont {H.}~\bibnamefont {Liu}}, \bibinfo {author} {\bibfnamefont {J.}~\bibnamefont {Martin}}, \ and\ \bibinfo {author} {\bibfnamefont {A.}~\bibnamefont {Roggero}},\ }\href {\doibase 10.1103/PhysRevResearch.7.023157} {\bibfield  {journal} {\bibinfo  {journal} {Phys. Rev. Res.}\ }\textbf {\bibinfo {volume} {7}},\ \bibinfo {pages} {023157} (\bibinfo {year} {2025})}\BibitemShut {NoStop}%
\bibitem [{\citenamefont {Martirosyan}\ \emph {et~al.}(2025)\citenamefont {Martirosyan}, \citenamefont {Gazo}, \citenamefont {Etrych}, \citenamefont {Fischer}, \citenamefont {Morris}, \citenamefont {Ho}, \citenamefont {Eigen},\ and\ \citenamefont {Hadzibabic}}]{coldatomspeed:2025}%
  \BibitemOpen
  \bibfield  {author} {\bibinfo {author} {\bibfnamefont {G.}~\bibnamefont {Martirosyan}}, \bibinfo {author} {\bibfnamefont {M.}~\bibnamefont {Gazo}}, \bibinfo {author} {\bibfnamefont {J.}~\bibnamefont {Etrych}}, \bibinfo {author} {\bibfnamefont {S.~M.}\ \bibnamefont {Fischer}}, \bibinfo {author} {\bibfnamefont {S.~J.}\ \bibnamefont {Morris}}, \bibinfo {author} {\bibfnamefont {C.~J.}\ \bibnamefont {Ho}}, \bibinfo {author} {\bibfnamefont {C.}~\bibnamefont {Eigen}}, \ and\ \bibinfo {author} {\bibfnamefont {Z.}~\bibnamefont {Hadzibabic}},\ }\href {\doibase 10.1038/s41586-025-09735-z} {\bibfield  {journal} {\bibinfo  {journal} {Nature}\ }\textbf {\bibinfo {volume} {647}},\ \bibinfo {pages} {608–612} (\bibinfo {year} {2025})}\BibitemShut {NoStop}%
\bibitem [{\citenamefont {Fisher}\ \emph {et~al.}(2023)\citenamefont {Fisher}, \citenamefont {Khemani}, \citenamefont {Nahum},\ and\ \citenamefont {Vijay}}]{Fisher-2023}%
  \BibitemOpen
  \bibfield  {author} {\bibinfo {author} {\bibfnamefont {M.~P.}\ \bibnamefont {Fisher}}, \bibinfo {author} {\bibfnamefont {V.}~\bibnamefont {Khemani}}, \bibinfo {author} {\bibfnamefont {A.}~\bibnamefont {Nahum}}, \ and\ \bibinfo {author} {\bibfnamefont {S.}~\bibnamefont {Vijay}},\ }\href {\doibase 10.1146/annurev-conmatphys-031720-030658} {\bibfield  {journal} {\bibinfo  {journal} {Annual Review of Condensed Matter Physics}\ }\textbf {\bibinfo {volume} {14}},\ \bibinfo {pages} {335–379} (\bibinfo {year} {2023})}\BibitemShut {NoStop}%
\bibitem [{\citenamefont {{Brown}}\ and\ \citenamefont {{Fawzi}}(2012)}]{Brown-2012}%
  \BibitemOpen
  \bibfield  {author} {\bibinfo {author} {\bibfnamefont {W.}~\bibnamefont {{Brown}}}\ and\ \bibinfo {author} {\bibfnamefont {O.}~\bibnamefont {{Fawzi}}},\ }\href {\doibase 10.48550/arXiv.1210.6644} {\bibfield  {journal} {\bibinfo  {journal} {arXiv e-prints}\ ,\ \bibinfo {eid} {arXiv:1210.6644}} (\bibinfo {year} {2012})},\ \Eprint {http://arxiv.org/abs/1210.6644} {arXiv:1210.6644 [quant-ph]} \BibitemShut {NoStop}%
\bibitem [{\citenamefont {Duan}\ \emph {et~al.}(2008)\citenamefont {Duan}, \citenamefont {Fuller}, \citenamefont {Carlson},\ and\ \citenamefont {Qian}}]{Duan-Nburst2008}%
  \BibitemOpen
  \bibfield  {author} {\bibinfo {author} {\bibfnamefont {H.}~\bibnamefont {Duan}}, \bibinfo {author} {\bibfnamefont {G.~M.}\ \bibnamefont {Fuller}}, \bibinfo {author} {\bibfnamefont {J.}~\bibnamefont {Carlson}}, \ and\ \bibinfo {author} {\bibfnamefont {Y.-Z.}\ \bibnamefont {Qian}},\ }\href {\doibase 10.1103/PhysRevLett.100.021101} {\bibfield  {journal} {\bibinfo  {journal} {Phys. Rev. Lett.}\ }\textbf {\bibinfo {volume} {100}},\ \bibinfo {pages} {021101} (\bibinfo {year} {2008})}\BibitemShut {NoStop}%
\bibitem [{\citenamefont {Qiu}\ \emph {et~al.}(2025)\citenamefont {Qiu}, \citenamefont {Radice}, \citenamefont {Richers},\ and\ \citenamefont {Bhattacharyya}}]{qiu:2025}%
  \BibitemOpen
  \bibfield  {author} {\bibinfo {author} {\bibfnamefont {Y.}~\bibnamefont {Qiu}}, \bibinfo {author} {\bibfnamefont {D.}~\bibnamefont {Radice}}, \bibinfo {author} {\bibfnamefont {S.}~\bibnamefont {Richers}}, \ and\ \bibinfo {author} {\bibfnamefont {M.}~\bibnamefont {Bhattacharyya}},\ }\href {\doibase 10.1103/h2q7-kn3v} {\bibfield  {journal} {\bibinfo  {journal} {Phys. Rev. Lett.}\ }\textbf {\bibinfo {volume} {135}},\ \bibinfo {pages} {091401} (\bibinfo {year} {2025})}\BibitemShut {NoStop}%
\bibitem [{\citenamefont {Wang}\ and\ \citenamefont {Burrows}(2025{\natexlab{a}})}]{wang2025effectfastflavor}%
  \BibitemOpen
  \bibfield  {author} {\bibinfo {author} {\bibfnamefont {T.}~\bibnamefont {Wang}}\ and\ \bibinfo {author} {\bibfnamefont {A.}~\bibnamefont {Burrows}},\ }\href {https://arxiv.org/abs/2503.04896} {\enquote {\bibinfo {title} {The effect of the fast-flavor instability on core-collapse supernova models},}\ } (\bibinfo {year} {2025}{\natexlab{a}}),\ \Eprint {http://arxiv.org/abs/2503.04896} {arXiv:2503.04896 [astro-ph.HE]} \BibitemShut {NoStop}%
\bibitem [{\citenamefont {Wang}\ \emph {et~al.}(2025)\citenamefont {Wang}, \citenamefont {Nagakura}, \citenamefont {Johns},\ and\ \citenamefont {Burrows}}]{wang2025effectcollisional}%
  \BibitemOpen
  \bibfield  {author} {\bibinfo {author} {\bibfnamefont {T.}~\bibnamefont {Wang}}, \bibinfo {author} {\bibfnamefont {H.}~\bibnamefont {Nagakura}}, \bibinfo {author} {\bibfnamefont {L.}~\bibnamefont {Johns}}, \ and\ \bibinfo {author} {\bibfnamefont {A.}~\bibnamefont {Burrows}},\ }\href {https://arxiv.org/abs/2507.01100} {\enquote {\bibinfo {title} {The effect of the collisional flavor instability on core-collapse supernova models},}\ } (\bibinfo {year} {2025}),\ \Eprint {http://arxiv.org/abs/2507.01100} {arXiv:2507.01100 [astro-ph.HE]} \BibitemShut {NoStop}%
\bibitem [{\citenamefont {Wang}\ and\ \citenamefont {Burrows}(2025{\natexlab{b}})}]{wang2025instabilitycorecollapse}%
  \BibitemOpen
  \bibfield  {author} {\bibinfo {author} {\bibfnamefont {T.}~\bibnamefont {Wang}}\ and\ \bibinfo {author} {\bibfnamefont {A.}~\bibnamefont {Burrows}},\ }\href {https://arxiv.org/abs/2511.20767} {\enquote {\bibinfo {title} {The effect of the fast-flavor instability on core-collapse supernova models: Ii. quasi-equipartition and the impact of various angular reconstruction methods},}\ } (\bibinfo {year} {2025}{\natexlab{b}}),\ \Eprint {http://arxiv.org/abs/2511.20767} {arXiv:2511.20767 [astro-ph.HE]} \BibitemShut {NoStop}%
\bibitem [{\citenamefont {Nachtmann}\ \emph {et~al.}()\citenamefont {Nachtmann} \emph {et~al.}}]{nachtmannelementary}%
  \BibitemOpen
  \bibfield  {author} {\bibinfo {author} {\bibfnamefont {O.}~\bibnamefont {Nachtmann}} \emph {et~al.},\ }\href@noop {} {\ }\BibitemShut {NoStop}%
\bibitem [{\citenamefont {Hagedorn}(1962)}]{hagedorn1962selected}%
  \BibitemOpen
  \bibfield  {author} {\bibinfo {author} {\bibfnamefont {R.}~\bibnamefont {Hagedorn}},\ }\href@noop {} {\emph {\bibinfo {title} {Selected topics on scattering theory: Part 1 Relativistic kinematics and precession of polarization}}},\ \bibinfo {type} {Tech. Rep.}\ (\bibinfo  {institution} {Cern},\ \bibinfo {year} {1962})\BibitemShut {NoStop}%
\end{thebibliography}

%

\end{document}